\documentclass[sigconf, nonacm]{acmart}
\usepackage{url}
\usepackage{balance}
\usepackage{listings}
\usepackage{xcolor}
\lstdefinestyle{sql}{
  language=SQL,
  basicstyle=\scriptsize\ttfamily,
  keywordstyle=\color[RGB]{236,0,124},
  stringstyle=\color[RGB]{153,0,25},
  commentstyle=\color[RGB]{51,128,89},
  morecomment=[l][\color{magenta}]{\#},
  morekeywords={SELECT, FROM, WHERE, INSERT, INTO, VALUES, UPDATE, SET, DELETE, JOIN, ON, PARTITION BY, PARTITION, OVER, ORDER BY},
  captionpos=b, 
  frame=single,                   
  tabsize=2,                      
  breaklines=true,                
    xleftmargin=3.4pt,
    xrightmargin=3.4pt,
}
\lstdefinestyle{info_queryplan}{
  language=C,
  basicstyle=\fontsize{6pt}{6pt}\ttfamily,
  keywordstyle=\color[RGB]{236,0,124},
  morecomment=[l][\color{magenta}]{\#},
  morekeywords={Filter, Sort, WindowAgg, IndexScan, Limit, SeqScan, Map},
    breaklines=true,                
}

\newcounter{sqlcount}
\newcommand{\customsqlcaption}[1]{
  \refstepcounter{sqlcount}
  \lstset{caption={\small\textmd{Q\thesqlcount: #1}}}
}

\usepackage{tabularx} 
\usepackage{pifont}   
\usepackage{multirow} 
\usepackage{array}    
\usepackage{threeparttable}

\usepackage{subcaption}
\usepackage{algorithm}
\usepackage{algpseudocode}

\usepackage{adjustbox}
\usepackage{tikz}

\lstdefinestyle{queryplan}{
  language=C,
  basicstyle=\scriptsize\ttfamily,
  keywordstyle=\color[RGB]{236,0,124},
  morekeywords={SortKey},     
}
\newcommand\vldbdoi{XX.XX/XXX.XX}
\newcommand\vldbpages{XXX-XXX}
\newcommand\vldbvolume{14}
\newcommand\vldbissue{1}
\newcommand\vldbyear{2020}
\newcommand\vldbauthors{\authors}
\newcommand\vldbtitle{\shorttitle} 
\newcommand\vldbavailabilityurl{URL_TO_YOUR_ARTIFACTS}
\newcommand\vldbpagestyle{plain} 

\begin{document}
\title{CHASE: A Native Relational Database for Hybrid Queries on Structured and Unstructured Data}

\author{Rui Ma}
\affiliation{%
  \institution{Fudan University}
}
\email{rma23@m.fudan.edu.cn}

\author{Kai Zhang}
\affiliation{%
  \institution{Fudan University}
}
\email{zhangk@fudan.edu.cn}

\author{Zhenying He}
\affiliation{%
  \institution{Fudan University}
}
\email{zhenying@fudan.edu.cn}

\author{Yinan Jing}
\affiliation{%
  \institution{Fudan University}
}
\email{jingyn@fudan.edu.cn}

\author{X.Sean Wang}
\affiliation{%
  \institution{Fudan University}
}
\email{xywangcs@fudan.edu.cn}

\author{Zhenqiang Chen}
\affiliation{%
  \institution{Transwarp}
}
\email{zhenqiang.chen@transwarp.io}

\begin{abstract}
Querying both structured and unstructured data has become a new paradigm in data analytics and recommendation. With unstructured data, such as text and videos, are converted to high-dimensional vectors and queried with approximate nearest neighbor search (ANNS). State-of-the-art database systems implement vector search as a plugin in the relational query engine, which tries to utilize the ANN index to enhance performance. After investigating a broad range of hybrid queries, we find that such designs may miss potential optimization opportunities and achieve suboptimal performance for certain queries.

In this paper, we propose CHASE, a query engine that is natively designed to support efficient hybrid queries on structured and unstructured data. CHASE performs specific designs and optimizations on multiple stages in query processing. First, semantic analysis is performed to categorize queries and optimize query plans dynamically. Second, new physical operators are implemented to avoid redundant computations, which is the case with existing operators. Third, compilation-based techniques are adopted for efficient machine code generation.
Extensive evaluations using real-world datasets demonstrate that CHASE achieves substantial performance improvements, with speedups ranging from 13\% to an extraordinary 7500\(\times\) compared to existing systems. These results highlight CHASE's potential as a robust solution for executing hybrid queries.


\end{abstract}

\maketitle

\pagestyle{\vldbpagestyle}
\begingroup\small\noindent\raggedright\textbf{PVLDB Reference Format:}\\
\vldbauthors. \vldbtitle. PVLDB, \vldbvolume(\vldbissue): \vldbpages, \vldbyear.\\
\href{https://doi.org/\vldbdoi}{doi:\vldbdoi}
\endgroup
\begingroup
\renewcommand\thefootnote{}\footnote{\noindent
This work is licensed under the Creative Commons BY-NC-ND 4.0 International License. Visit \url{https://creativecommons.org/licenses/by-nc-nd/4.0/} to view a copy of this license. For any use beyond those covered by this license, obtain permission by emailing \href{mailto:info@vldb.org}{info@vldb.org}. Copyright is held by the owner/author(s). Publication rights licensed to the VLDB Endowment. \\
\raggedright Proceedings of the VLDB Endowment, Vol. \vldbvolume, No. \vldbissue\ %
ISSN 2150-8097. \\
\href{https://doi.org/\vldbdoi}{doi:\vldbdoi} \\
}\addtocounter{footnote}{-1}\endgroup

\ifdefempty{\vldbavailabilityurl}{}{
\vspace{.3cm}
\begingroup\small\noindent\raggedright\textbf{PVLDB Artifact Availability:}\\
The source code, data, and/or other artifacts have been made available at \url{\vldbavailabilityurl}.
\endgroup
}

\section{Introduction}
In modern applications such as recommendation systems\cite{zhang2023vbase, wang2024efficient}, image retrieval\cite{bigann2023, yang2021cross, wen2023target}, and e-commerce \cite{wei2020analyticdb, wang2021milvus, patel2024acorn}, users often perform hybrid queries on both unstructured and structured data for richer search capabilities\cite{patel2024lotus, biswal2024text2sql, maddendatabases, zhang2024there}. For example, e-commerce users may search for products similar to a reference image while filtering by price\cite{wei2020analyticdb}, recipe systems may retrieve dishes matching specified ingredients and image similarity\cite{zhang2023vbase}, and image retrieval applications may find pictures matching specific landscape features and tagged with certain shooting years\cite{bigann2023}. Queries on structured data have been widely studied in database systems, which rely on well-defined schemas to perform specific filtering, sorting, and aggregation conditions to obtain precise results. On the other hand, unstructured data, such as images, texts, and videos, are typically transformed into high-dimensional embeddings for effective similarity searches, balancing precision and performance. When multimodal applications exhibit a high demand for hybrid queries, modern database systems should evolve to perform efficient execution. Currently, there is a lack of systematic studies on query engines to provide efficient support for such queries.

Relational databases are highly efficient for querying structured data, where query processing optimizes performance through multiple stages, including query parsing, query rewriting, and cost estimation, to select physical operators. In querying unstructured data, the vectors are generally taken as another column in the relational table. 
Since relational databases generally provide developers with customizable index interfaces, considerable work has incorporated the ANN index into relational databases to support hybrid queries, such as PASE\cite{yang2020pase}, AnalyticDB-V (ADBV)\cite{wei2020analyticdb}, pgvector\cite{pgvector}, and VBASE\cite{zhang2023vbase}. With built indices, ANN search \cite{jayaram2019diskann, malkov2018efficient, fu2017fast} is used to perform similarity searches, just as the search on a B+ tree for structured attributes. Therefore, the query engine in a relational database can leverage ANN indices in the \textit{index scan} operator on vectors, which can significantly enhance the performance of hybrid search.

\begin{figure*}[h!]
    \centering
    \begin{minipage}{0.34\textwidth}
        \centering
        \begin{lstlisting}[
        style=sql,
        basicstyle=\fontsize{6pt}{6pt}\ttfamily,
        frame=none,
        xleftmargin=0pt,
        xrightmargin=0pt
        ]
SELECT id, embedding 
FROM products 
WHERE price < 100
ORDER BY DISTANCE(embedding, ${ image_embedding })
LIMIT 50;
        \end{lstlisting}
        \subcaption{Hybrid query example}
        \label{fig:exampleknn}
        \vspace{2pt}
        \begin{lstlisting}[style=info_queryplan]
Limit (rows=50)
->  Filter [price < 100]
->  IndexScan on products [Order By: vec <*> query]
        \end{lstlisting}
        \subcaption{The query plan of PASE}
        \label{fig:exampleknn-plan-a}
    \end{minipage}
    \hfill
    \hspace{2pt}
    \begin{minipage}{0.34\textwidth}
        \centering
        \begin{lstlisting}[style=info_queryplan]
Limit (rows=50)
->  Sort [Key: vec <*> query]
->  Filter [price < 100]
->  IndexScan on products [Order By: vec <*> query]
        \end{lstlisting}
        \subcaption{The query plan of VBASE}
        \label{fig:exampleknn-plan-b}
        \vspace{2pt}
        \begin{lstlisting}[style=info_queryplan]
Limit (rows=50)
->  Sort [Key: sim]
->  Filter [price < 100]
->  Map [sim: vec <*> query]
->  IndexScan on products [Order By: vec <*> query]
        \end{lstlisting}
        \subcaption{The query plan of CHASE}
        \label{fig:exampleknn-plan-c}
    \end{minipage}
    \hfill
    \begin{minipage}{0.29\textwidth}
        \centering
        \includegraphics[width=0.7\textwidth]{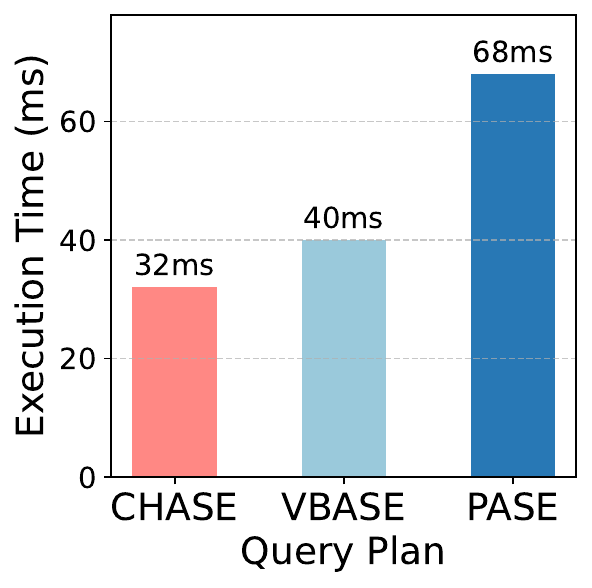}
        \subcaption{Performance comparison for query plan of PASE, VBASE and CHASE}
        \label{fig:intromap}
    \end{minipage}
    \caption{Hybrid query and query qlans}
    \label{fig:hybrid-query-example}
\end{figure*}

Through query plan analysis and performance comparison, we find that the current approach in databases has inherent limitations to support hybrid queries. This is because, although leveraging ANN search can optimize the performance of the \textit{scan} operator, the influences on other operators in the hybrid query plan are neglected.
For instance, the hybrid query in Figure \ref{fig:exampleknn} retrieves similar products from the "products" table according to an image, where the price is below 100. The similarity is ranked by the DISTANCE function on embedding vectors. Figure \ref{fig:exampleknn-plan-a} shows a query plan where an \textit{index scan} operator utilizes an ANN index with the distance function (e.g., \textit{vec <*> query}) to retrieve tuples, applying filters on attributes. This optimization accelerates the data retrieval process by avoiding performing distance comparisons on all vectors whose attributes satisfy the filters. However, systems adopting this approach, such as PASE, pgvector, and ADBV, often use a conservatively large \( K' \) (\( K' \gg K \)), where \( K \) denotes the desired number of results (e.g., top 50 most similar products), to ensure sufficient results meet the constraints, resulting in substantial redundant computations.
Figure \ref{fig:exampleknn-plan-b} illustrates another query plan generated by the VBASE system, where the query optimizer also uses an \textit{index scan} operator that leverages the ANN index to retrieve tuples similar to the query vector. However, there are repetitive similarity computations on the vectors in the ANN-based \textit{index scan} operator and the subsequent \textit{sort} operator. Overall, we find that both query plans lack sufficient optimizations, resulting in suboptimal performance.
The key reason for the inefficient query plans is that existing databases do not take the vector attribute as a first-class citizen, and there is a lack of specific design and optimizations for hybrid queries.

We propose CHASE, a query engine that is natively designed to
support efficient hybrid queries on structured and unstructured data. CHASE is designed to comprehensively optimize hybrid query execution by integrating optimization techniques across all stages of the query processing pipeline, including logical plan optimization, physical operator optimization, and machine code generation. 
First, in logical plan optimization, CHASE conducts a semantic analysis to identify the type of hybrid query and rewrites the logical plan to reduce redundant computations. 
For example, in hybrid queries like the one depicted in Figure \ref{fig:exampleknn}, CHASE implements a \textit{map} operator to extract computed similarity results from index scans and map them to a temporary column. This column is then directly used by sorting operators, thereby avoiding unnecessary recomputations. 
Next, in physical operator optimization, CHASE improves query execution efficiency by optimizing the implementation of physical operators based on the specific characteristics of hybrid queries.
Taking the hybrid query shown in Figure \ref{fig:exampleknn} as an example, CHASE optimizes the \textit{scan} operator so that, after traversing the ANN index, the \textit{scan} operator not only returns the tuples found but also the similarity scores computed during the index traversal. These scores are then utilized by the \textit{map} operator introduced in the logical plan optimization phase. The modified query plan, which eliminates repetitive computations in both the \textit{Sort} and \textit{Scan} phases, is illustrated in Figure \ref{fig:exampleknn-plan-c}. By optimizing the logical and physical plans to remove redundant computations, the execution time for the hybrid query is reduced compared to the query plan of PASE and VBASE. As shown in Figure \ref{fig:intromap}, CHASE achieves an execution time of 32 ms, outperforming VBASE (40 ms) and PASE (68 ms).
Finally, CHASE generates machine code to compile optimized query plans, executing queries directly on hardware. This eliminates interpretation overhead and enables low-level optimizations, such as instruction pipelining, branch prediction, and memory access improvements, maximizing processor efficiency.

To validate the effectiveness of our approach, we conduct extensive evaluations using real-world datasets. We analyze a broad class of hybrid queries in modern applications, including top-k hybrid queries, distance-based range hybrid queries, distance-join hybrid queries, KNN-join hybrid queries, category-based partition hybrid queries, and category-join hybrid queries.
We analyze their overhead and propose corresponding optimizations in CHASE.
The evaluation results show that CHASE significantly enhances performance and balances accuracy. Compared to existing systems, CHASE achieves performance gains ranging from 17\% to 7,500$\times$, including up to a 33\% improvement for top-k queries, 24\% to 33\% for distance-based range queries, approximately 64\% for distance-join queries, 7,500$\times$ for KNN-join queries, 33\% to 46\% for category-based partition queries, and 3.13$\times$ to 4.04$\times$ for category-join queries.

\section{Background and Motivavtion} \label{sec:hybrid_query}
\begin{figure*}[ht]
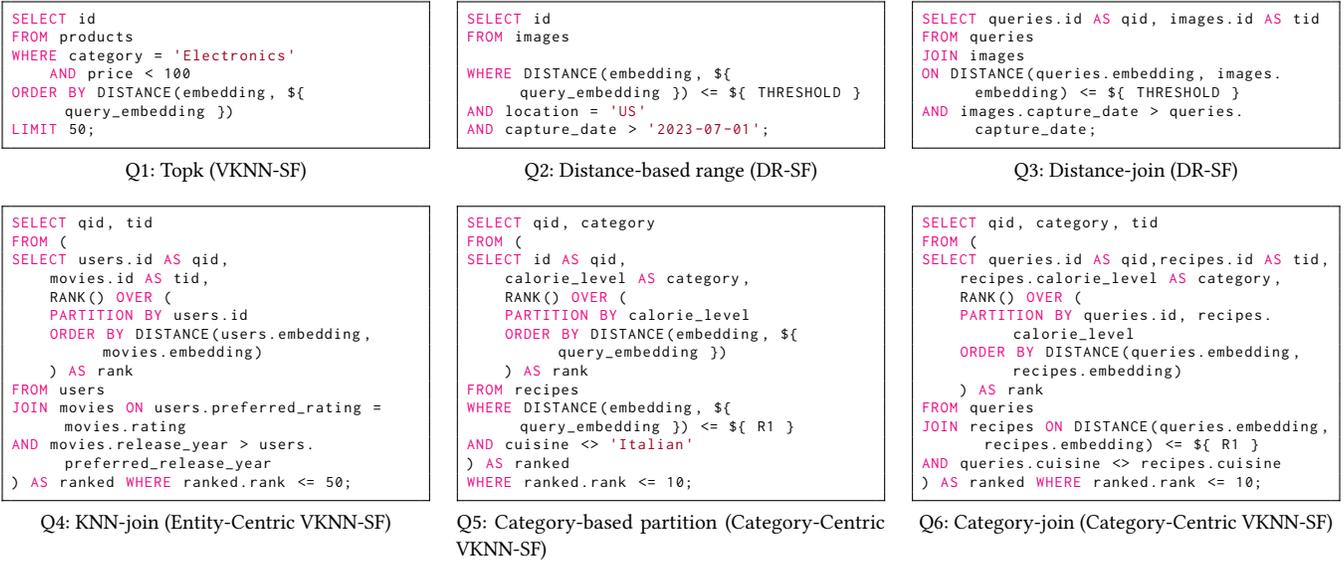

    \centering
    \begin{minipage}[t]{0.32\textwidth}
\customsqlcaption{Topk (VKNN-SF)}
\begin{lstlisting}[style=sql]
SELECT id
FROM products
WHERE category = 'Electronics' 
    AND price < 100
ORDER BY DISTANCE(embedding, ${ query_embedding })
LIMIT 50;
\end{lstlisting}
    \end{minipage}\hfill
    \begin{minipage}[t]{0.32\textwidth}
\customsqlcaption{Distance-based range (DR-SF)}
\begin{lstlisting}[style=sql]
SELECT id
FROM images

WHERE DISTANCE(embedding, ${ query_embedding }) <= ${ THRESHOLD } 
AND location = 'US'
AND capture_date > '2023-07-01';
\end{lstlisting}
    \end{minipage}\hfill
    \begin{minipage}[t]{0.32\textwidth}
\customsqlcaption{Distance-join (DR-SF)}
\begin{lstlisting}[style=sql]
SELECT queries.id AS qid, images.id AS tid 
FROM queries 
JOIN images
ON DISTANCE(queries.embedding, images.embedding) <= ${ THRESHOLD } 
AND images.capture_date > queries.capture_date;
\end{lstlisting}
    \end{minipage}
    
    \begin{minipage}[t]{0.32\textwidth}
\customsqlcaption{KNN-join (Entity-Centric VKNN-SF)}
\begin{lstlisting}[style=sql]
SELECT qid, tid
FROM (
SELECT users.id AS qid, 
    movies.id AS tid,
    RANK() OVER (
    PARTITION BY users.id
    ORDER BY DISTANCE(users.embedding, movies.embedding)
    ) AS rank
FROM users
JOIN movies ON users.preferred_rating = movies.rating 
AND movies.release_year > users.preferred_release_year
) AS ranked WHERE ranked.rank <= 50;
\end{lstlisting}
    \end{minipage}\hfill
    \begin{minipage}[t]{0.32\textwidth}
\customsqlcaption{Category-based partition (Category-Centric VKNN-SF)}
\begin{lstlisting}[style=sql]
SELECT qid, category
FROM (
SELECT id AS qid, 
    calorie_level AS category, 
    RANK() OVER (
    PARTITION BY calorie_level
    ORDER BY DISTANCE(embedding, ${ query_embedding })
    ) AS rank
FROM recipes
WHERE DISTANCE(embedding, ${ query_embedding }) <= ${ R1 } 
AND cuisine <> 'Italian'
) AS ranked
WHERE ranked.rank <= 10;
\end{lstlisting}
    \end{minipage}\hfill
    \begin{minipage}[t]{0.32\textwidth}
\customsqlcaption{Category-join (Category-Centric VKNN-SF)}
\begin{lstlisting}[style=sql]
SELECT qid, category, tid
FROM (
SELECT queries.id AS qid,recipes.id AS tid, 
    recipes.calorie_level AS category,
    RANK() OVER (
    PARTITION BY queries.id, recipes.calorie_level
    ORDER BY DISTANCE(queries.embedding, recipes.embedding)
    ) AS rank
FROM queries 
JOIN recipes ON DISTANCE(queries.embedding, recipes.embedding) <= ${ R1 } 
AND queries.cuisine <> recipes.cuisine 
) AS ranked WHERE ranked.rank <= 10;
\end{lstlisting}
    \end{minipage}
    \caption{Hybrid query examples}
    \label{fig:sql_code_listings}
\end{figure*}

\subsection{Background and Advances in Hybrid Queries Processing}
Unstructured data, such as text, images, and videos, contains rich semantic and contextual information but presents challenges for traditional data processing techniques. To address these, advanced techniques, such as deep learning models, transform unstructured data into vector representations, capturing latent features and relationships in high-dimensional spaces \cite{chen2021learning, jia2021scaling, radford2021learning, shi2020towards}. This transformation enables efficient semantic querying using vector similarity search \cite{zheng2024adapting, lin2024parrot, dai2019deeper, hu2024avis, lai2024lisa}. As the number and dimensions of feature vectors increase, traversing the entire database to complete a semantic query becomes infeasible. To enhance retrieval efficiency, ANN algorithms have emerged\cite{lu2022mqh, tian2023db, li2023learning, aguerrebere2023similarity, paparrizos2022fast}. ANN algorithms enable the rapid discovery of vectors similar to the query vector within large datasets by constructing vector indices, allowing for a significant improvement in search speed while tolerating minor accuracy losses. 

Modern applications increasingly demand semantic queries that not only rely on vector similarity search but also incorporate structured data for more precise control. For instance, when recommending a lightweight backpack, users may also need to filter results by specific attributes, such as price range, user ratings, or stock availability. Through a comprehensive survey across various fields such as recommendation systems\cite{lee2021bootstrapping, kweon2021bidirectional, kweon2024top, chen2021multi, wu2024result, cen2020controllable}, image retrieval \cite{jia2023unitsface, wen2023target, yang2021cross, schroff2015facenet, hosseinzadeh2020composed}, and information retrieval \cite{lewis2020retrieval, wang2021milvus, wang2024efficient, maddendatabases}, we identify and summarize three common types of hybrid queries: 1) Vector KNN with Structured Data Filter (VKNN-SF) queries, 2) Distance-based Range with Structured Data Filter (DR-SF) queries, and 3) Window Vector KNN with Structured Data Filter (W-VKNN-SF) queries. These queries are widely encountered across diverse application areas, each addressing distinct needs for combining vector-based similarity with structured data constraints.

To address the dual demands of vector similarity retrieval and structured data filtering in hybrid queries, relational databases have emerged as a promising solution. Their ability to efficiently manage structured data, coupled with inherent support for indexing mechanisms, makes them well-suited for integrating ANN indices to handle hybrid queries. Systems such as PASE, AnalyticDB-V (ADBV), pgvector, and VBASE exemplify this integration, blending vector similarity search with relational operations to deliver a unified query execution engine. These systems enhance hybrid query processing by incorporating ANN indices into the database's index interface, enabling the physical plan generation process to replace logical \textit{scan} operator with \textit{index scan} operator, thereby leveraging ANN indices to accelerate the data scanning phase. 
However, our analysis reveals that optimizations focusing solely on accelerating the scan phase lead to notable limitations in existing systems when handling hybrid queries. For instance, all current systems exhibit redundant computations in processing VKNN-SF queries. Furthermore, most systems, such as PASE, pgvector, and ADBV, are unable to leverage ANN techniques to process all types of DR-SF and W-VKNN-SF queries. Similarly, while VBASE achieves optimizations in certain W-VKNN-SF scenarios, it still suffers from redundant computations and is unable to utilize ANN indices for some query cases effectively.
These issues stem from the lack of a comprehensive analysis of query plans for hybrid queries, highlighting the need for deeper optimizations that extend beyond the scan phase to address such challenges effectively. In the subsequent discussion, we will examine how these limitations influence the overall query plan and propose strategies to address them.  

\subsection{Vector KNN with Structured Data Filter Queries} 
VKNN-SF queries combine both similarity measurements and structured data filters to retrieve the top \( K \) elements most similar to a given query vector. These queries are particularly valuable in applications where precision and contextual relevance are critical, such as recommendation systems \cite{lee2021bootstrapping, kweon2021bidirectional, kweon2024top}, retrieval-augmented generation (RAG) \cite{lewis2020retrieval}, and machine translation \cite{agrawal2022context}. 
By integrating structured data with vector similarity, VKNN-SF queries enhance precision and contextual relevance, providing more accurate results in domains like product recommendations, content retrieval, and language translation \cite{wang2024efficient, maddendatabases, yang2020pase, wei2020analyticdb, zhang2024there, wang2021milvus}.
Formally, this can be expressed as:
\[
\text{VKNN-SF}(q, D, K, F) = \underset{x \in D \land F(x)}{\text{argTopK}} \ \text{distance}(x, q)
\]
where $\text{distance}(x, q)$ denotes the similarity between vectors $x$ and $q$, and $F(x)$ represents the filtering criteria based on structured data attributes, such as time range, category, or other relational constraints. VKNN-SF queries can be expressed in an SQL format similar to Q1 in Figure \ref{fig:sql_code_listings}. The database system first applies the structured conditions from the WHERE clause. Then, it calculates the similarity between each row's embedding and the query's embedding using a DISTANCE function. The rows are sorted by similarity, and the LIMIT clause ensures that only the top 50 results are returned. 

There are two prevalent approaches to executing Q1. 
The first approach, implemented in systems such as PASE, pgvector, and ADBV, often requires conservatively selecting a large \( K' \) value (\( K' \gg K \)) to ensure that \( K \) results satisfying the structural constraints can be retrieved. This strategy, however, leads to substantial redundant computations.
The second approach, proposed in VBASE, introduces the concept of relaxed monotonicity by retrieving one tuple at a time rather than fetching \( K' \) tuples in a single batch. This incremental processing method enables the database to dynamically adjust the query execution process, continuing until the desired \( K \) relevant tuples are obtained, thereby avoiding the retrieval of excessive redundant data. However, as illustrated in Figure \ref{fig:exampleknn-plan-b}, this approach transforms the original strict monotonicity into relaxed monotonicity, necessitating the use of a \textit{sort} operator to ensure strictly monotonic results. Consequently, both the \textit{index scan} operator and the \textit{sort} operator redundantly compute similarity for the same vector pairs (\textit{vec <*> query}), leading to increased computational overhead.

\subsection{Distance-based Range with Structured Data Filter Queries}
DR-SF queries incorporate both similarity measures and filtering conditions to retrieve all elements whose distance is below a given threshold, which is equivalent to selecting those whose similarity exceeds a specified value. Unlike VKNN-SF queries, which return a fixed number of results, DR-SF queries offer more flexibility by returning all elements that meet the similarity threshold. These queries have significant applications in image retrieval tasks, where they enable the identification of images that closely match a target by combining vector similarity with additional metadata filters such as time, location, or category \cite{jia2023unitsface, hosseinzadeh2020composed, schroff2015facenet, yang2021cross, wen2023target}.
Formally, given a query vector $q$, a dataset $D$, a similarity threshold $\epsilon$, and a filter function $F$, the query can be defined as:
\[
\text{DR-SF}(q, D, \epsilon, F) = \{ x \in D \mid \text{distance}(x, q) \leq \epsilon \land F(x) \}
\]
The SQL representation for this type of query is provided in Figure \ref{fig:sql_code_listings} for Q2 and Q3, where Q3 illustrates a join-based extension of Q2. Taking Q2 as an example, the database system first evaluates the conditions in the WHERE clause to filter the records, including both structured conditions and the unstructured condition, where the DISTANCE function computes the distance between each row’s embedding and the provided \(query\_embedding\). The results are then selected based on distances that are less than the specified threshold. 

In the queries of Q2 and Q3, since the distance function in the WHERE clause corresponds to an indexed vector column, the system could, in theory, leverage the ANN index to process the query, in accordance with the optimization rules of relational databases. However, some systems, such as PASE, pgvector, and ADBV, while supporting ANN indices, are limited to efficiently retrieving the \(K\) nearest neighbors. This limitation prevents the use of ANN indices for processing DR-SF queries, necessitating the use of brute-force search instead.
\subsection{Window Vector KNN with Structured Data Filter Queries}
W-VKNN-SF queries partition the dataset into subsets based on a specific key (e.g., user-defined classes or clusters), and KNN queries are executed independently within each subset. The results from all subsets are then combined. In contrast to the previous two hybrid types of queries, the goal of W-VKNN-SF queries is to identify the top \( K \) most similar items within each partition, allowing for diverse recommendations that span different categories. These queries are particularly useful in recommendation systems, where data can be partitioned into categories or subsets, such as movie genres or user clusters \cite{cen2020controllable, wu2024result, chen2021multi}. 
Let $C_i$ represent a subset of $D$, where elements are grouped by a specific key (e.g., user ID or calorie level of the recipe). For each subset $C_i$, the top $K$ nearest neighbors are found as follows:
\[
\text{W-VKNN-SF}(q, D, K, F) = \bigcup_{i=1}^n \left( \text{VKNN-SF}(q, C_i, K, F) \mid C_i \subseteq D \right)
\]
W-VKNN-SF queries can be categorized into two distinct scenarios.

\textbf{Entity-Centric VKNN-SF Queries. }The first scenario, as illustrated by Q4 in Figure \ref{fig:sql_code_listings}, involves identifying the K most similar videos in the movies table for each user, thereby recommending relevant content. The Entity-Centric VKNN-SF queries partition the dataset by the primary key to construct subsets \(C_i\), where each partition \(C_i\) contains the records associated with a specific primary key value, ensuring that each entity (e.g., user) is assigned to a distinct partition for similarity calculations and ranking among its associated records (e.g., movies).

When executing Q4, existing systems generate a logical plan, as shown in Figure \ref{fig:unknnjoin_lplan}. In this logical plan, the vector field \textit{B.vec} referenced in the ORDER BY clause is associated with an ANN index, which, in theory, allows the system to accelerate scan phase through index-based similarity searches. However, the presence of a PARTITION BY clause in the \textit{window} operator adds the \textit{id} field as the primary sorting key, disrupting the intended order and preventing efficient index utilization. Consequently, the query engine falls back to brute-force processing, significantly increasing the execution time to \( O(|A| \cdot |B| + |A| \cdot |B| \cdot \log |B|) \), where \( |A| \) represents the number of tuples in the "users" table and \( |B| \) represents the number of tuples in the "movies" table. Ideally, the ANN index would perform VKNN-SF queries for each record in table "users", reducing the execution time to \(O(C \cdot |A|)\), where \(C\) is the cost of similarity comparisons using the ANN index. It is evident that the practical execution time of Q4 is significantly higher than in the ideal scenario.

\textbf{Category-Driven VKNN-SF Queries. }In the second scenario, as illustrated by Q5 and Q6 in Figure \ref{fig:sql_code_listings}, the system first identifies a set of recipes within a specified similarity threshold to the user's preferences, which span across various recipe categories (e.g., low-calorie, high-calorie). Then, for each calorie level within this set, the top \(K\) most similar recipes are selected. Q6 can be viewed as a join-based extension of Q5, where the same process is applied, but with the added complexity of a join operation. 
The Category-Driven VKNN-SF queries partition the dataset not only by the primary key of one table but also by an additional categorical field from the other table, constructing subsets \(C_i\) for each distinct combination of the primary key and category. For Q5, this can be viewed as having a constant primary key, such as (PARTITION BY 1, category). 
Compared to the Entity-Centric VKNN-SF query, this query ensures that each entity corresponding to a unique primary key is assigned to a distinct partition, while further partitioning the data by the categorical dimension. This allows the relevance ranking of each entity to be performed separately within multiple categories, which is crucial for ensuring both accuracy and diversity in recommendation systems \cite{cen2020controllable, wu2024result, chen2021multi}. 

For queries Q5 and Q6, the goal is to categorize data within the range \( R_1 \), centered around the query vector, and identify the \( K \) nearest neighbors for each category. As \( R_1 \) increases, the amount of data to be processed grows. 
Assuming that the number of tuples per category within \( R_1 \) is \( K' \), where \( K' \geq K \), and categories are uniformly distributed, then there exists a smaller range \( R_2 \) in which reducing the range \( R_1 \) to \( R_2 \) still allows for identifying the \( K \) nearest neighbors per category. This is because, under the assumption of uniform distribution, reducing the range does not reduce the number of categories within the query range. Furthermore, since the number of tuples per category within \( R_1 \) is greater than \( K \), reducing the range to \( R_2 \) still ensures that \( K \) nearest neighbors can be identified for each category.
In other words, reducing the tuples to be traversed from \( S(R_1) \) to \( S(R_2) \) does not affect the query results. However, existing systems that support Category-Driven VKNN-SF queries can only retrieve records within \( R_1 \) and cannot dynamically shrink the search range to \( R_2 \), leading to unnecessary computational overhead.

Moreover, most works handle all hybrid queries by generating a physical plan based on a cost model, subsequently selecting and executing the physical plan with the lowest overall cost. The execution process of these databases typically employs an iterator model in which the root node repeatedly invokes the \textit{Next} function to retrieve results generated by its child nodes.
While the iterator model is conceptually simple, it incurs substantial performance overhead due to the frequent invocation of the \textit{Next} function to generate tuples\cite{neumann2011efficiently}. Additionally, these function calls are typically implemented via virtual calls or function pointers, where the target address of the function remains unresolved at compile time\cite{neumann2011efficiently}. This ambiguity complicates the CPU’s branch prediction mechanism, and each misprediction forces a pipeline flush, significantly increasing instruction execution latency.

In conclusion, although existing relational databases leverage ANN indices to accelerate data scans and enhance query performance, relying solely on optimizing the scan phase is insufficient to fully optimize the execution process for hybrid queries.

\section{Overview} \label{sec:overview}
\begin{figure}[h]
    \centering
    \includegraphics[width=0.46\textwidth]{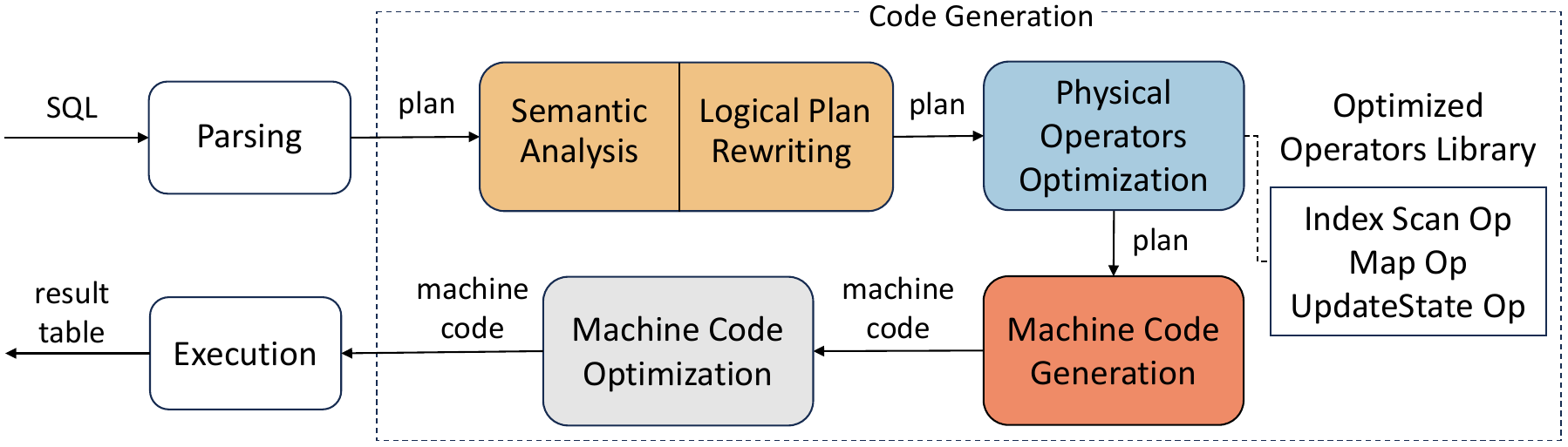}
    \caption{Architecture for hybrid queries processing}
    \label{fig:system_overview}
\end{figure}

The general query processing workflow of traditional relational databases fails to fully address the complexity of hybrid queries. Essentially, databases treat vectors as ordinary relational data, only selecting ANN indices based on query conditions to accelerate the scan phase. However, this optimization strategy fails to resolve the issues identified in Section \ref{sec:hybrid_query}, which limits their performance in handling hybrid queries.
To address the challenges associated with hybrid queries, we propose CHASE, a new database system that extends the traditional SQL query processing workflow of relational databases. CHASE improves execution efficiency by optimizing hybrid query execution at multiple levels, including logical plan optimization, physical operator optimization, and machine code generation.

First, as shown in Figure \ref{fig:system_overview}, the system parses the SQL into an initial query plan. Subsequently, the system performs semantic analysis to identify the type of hybrid query and determine whether the query can leverage ANN indices to accelerate hybrid query processing. Based on this analysis, the system rewrites the logical plan to minimize computational overhead at the logical level. For instance, to avoid redundant similarity computations in Q1, a new logical operator is introduced. This operator extracts computed similarity results from the \textit{scan} operator and maps them to a temporary column, which can then be directly utilized by the \textit{orderBy} operator. For Q4, the \textit{orderBy} operator within the \textit{window} operator can be decoupled, and an explicit \textit{limit} operator can be added to enable the query to leverage ANN index-based processing. In the case of Q5 and Q6, a new logical operator is introduced to dynamically narrow the query range, enabling the \textit{scan} operator to terminate the traversal of the ANN index early.

Secondly, CHASE implements an optimized operator library for hybrid queries, from which the framework selects the appropriate physical operators for logical operators. For example, the \textit{index scan} operator for Q2 and Q3 should utilize the RangeSearch interface from the ANN index, rather than the Topk interface. This allows the query to leverage the ANN index efficiently, eliminating the need for a full table scan and significantly reducing computational overhead. Similarly, for the newly introduced logical operators added during the logical plan rewriting process, CHASE assigns optimized physical operators to ensure efficient query execution.
    
Finally, the optimized query plan is compiled into machine code, allowing for more efficient execution by directly leveraging the hardware's capabilities. This compilation step eliminates the overhead associated with query interpretation and enables the use of low-level optimizations, such as instruction pipelining, branch prediction, and efficient memory access patterns. By generating machine code tailored to the specific hardware architecture, CHASE can exploit the full potential of modern processors, significantly reducing hybrid query execution time.  

\begin{table}[ht]
\centering
\caption{Query optimizating stages for Q1-Q6}
\begin{tabular}{|>{\centering}m{34mm}|c|c|c|c|c|c|}
\hline
Stage & Q1 & Q2 & Q3 & Q4 & Q5 & Q6 \\
\hline
Logical Plan Rewriting & \checkmark &  &  & \checkmark & \checkmark & \checkmark \\
\hline
Physical Operator Optimization & \checkmark & \checkmark & \checkmark & \checkmark & \checkmark & \checkmark \\
\hline
Machine Code Generation & \checkmark & \checkmark & \checkmark & \checkmark & \checkmark & \checkmark \\
\hline
\end{tabular}
\label{tab:query_processing_stages}
\end{table}

Table \ref{tab:query_processing_stages} summarizes the specific stages where different types of hybrid queries (Q1-Q6) benefit from CHASE's optimizations. 
Through this end-to-end optimization process, CHASE significantly enhances the execution efficiency of hybrid queries.

\section{Logical Plan Rewriting} \label{sec:query_rewriting}

The query engine performs semantic analysis by traversing the logical plan to ensure that the logical plan of the hybrid query conforms to the corresponding pattern, which guarantees alignment with the semantics of a specific category of hybrid queries, and then rewrites the logical plan accordingly.
\begin{figure}[ht]
    \centering
    \begin{subfigure}[b]{0.21\textwidth}
    \captionsetup{font=scriptsize}
         \centering
         \includegraphics[width=0.7\textwidth]{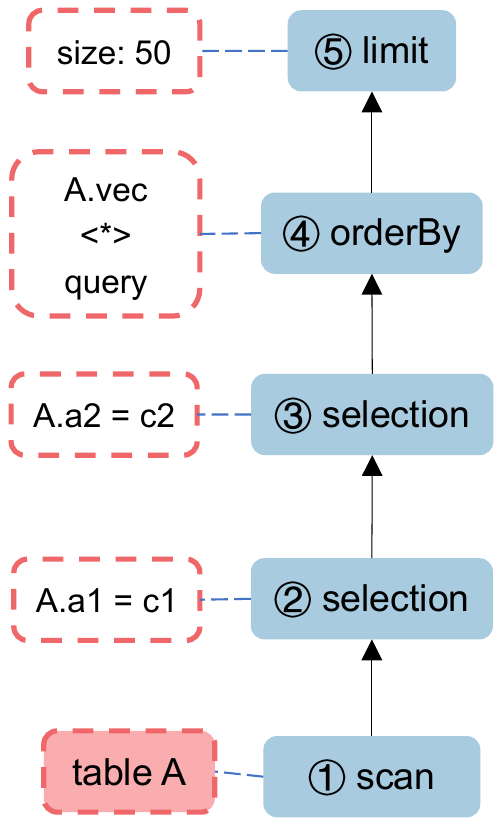}
         \caption{Unoptimized logical plan for Q1}
         \label{fig:untopk_lplan}
     \end{subfigure}
     \begin{subfigure}[b]{0.21\textwidth}
    \captionsetup{font=scriptsize}
         \centering
         \includegraphics[width=0.78\textwidth]{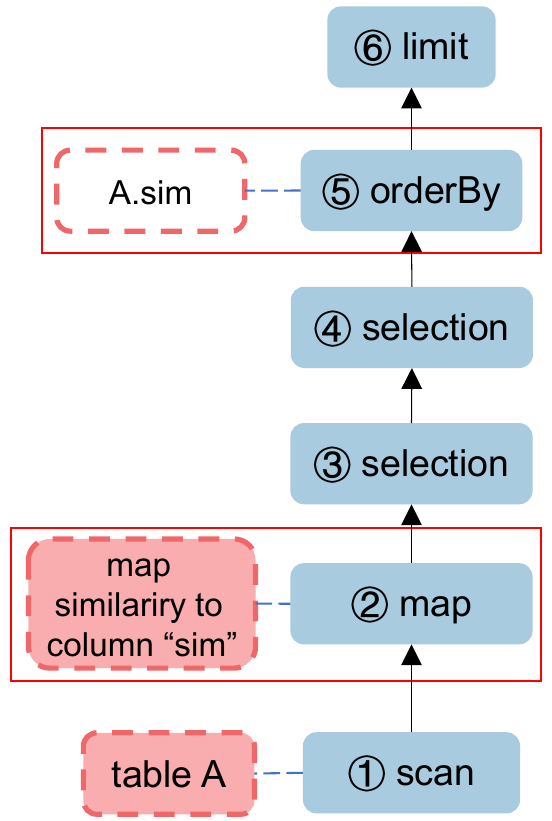}
        \caption{Optimized logical plan for Q1}
         \label{fig:topk_lplan}
     \end{subfigure}
    \caption{Logical plan comparison for Q1}
\end{figure}
\subsection{Rewriting KNN-like Queries}
\begin{figure}[ht]
    \centering
    \begin{subfigure}[b]{0.23\textwidth}
    \captionsetup{font=scriptsize}
         \centering
         \includegraphics[width=1\textwidth]{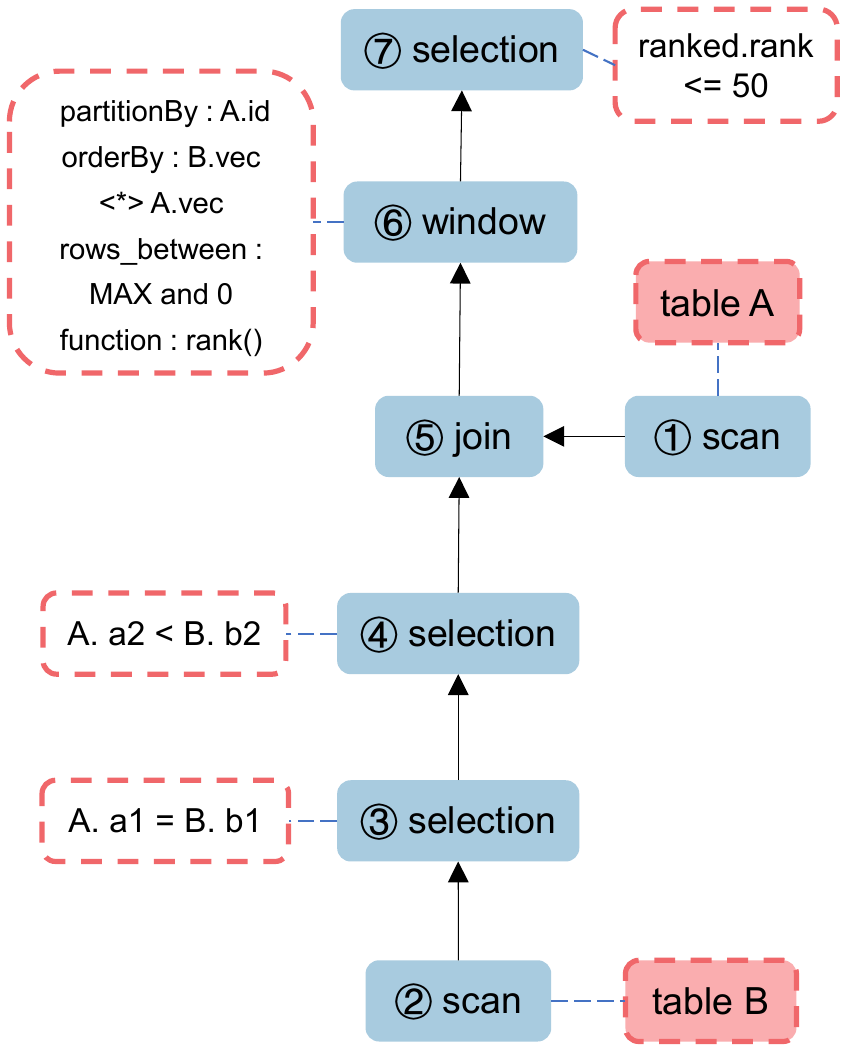}
         \caption{Unoptimized logical plan for Q4}
         \label{fig:unknnjoin_lplan}
     \end{subfigure}
     \begin{subfigure}[b]{0.23\textwidth}
    \captionsetup{font=scriptsize}
         \centering
         \includegraphics[width=1\textwidth]{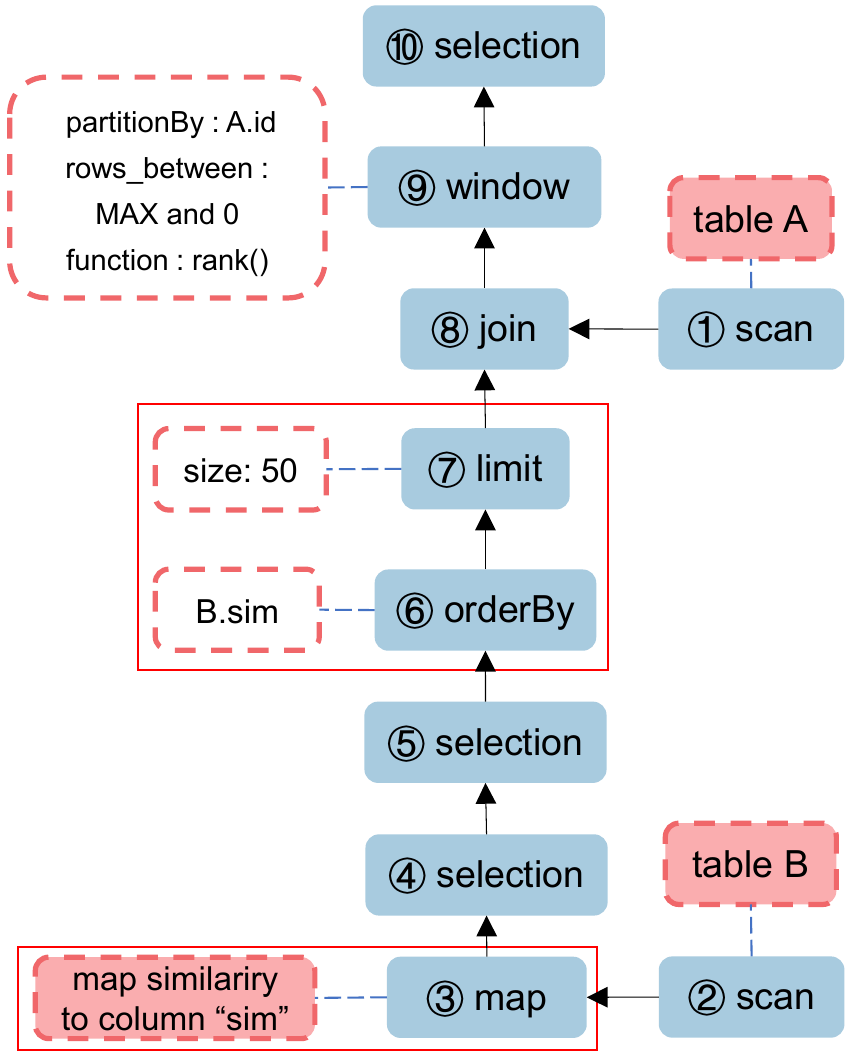}
        \caption{Optimized logical plan for Q4}
         \label{fig:knnjoin_lplan}
     \end{subfigure}
    \caption{Logical plan comparison for Q4}
\end{figure}

\begin{figure}[ht]
    \centering
    \begin{subfigure}[b]{0.2321\textwidth}
    \captionsetup{font=scriptsize}
         \centering
         \includegraphics[width=1\textwidth]{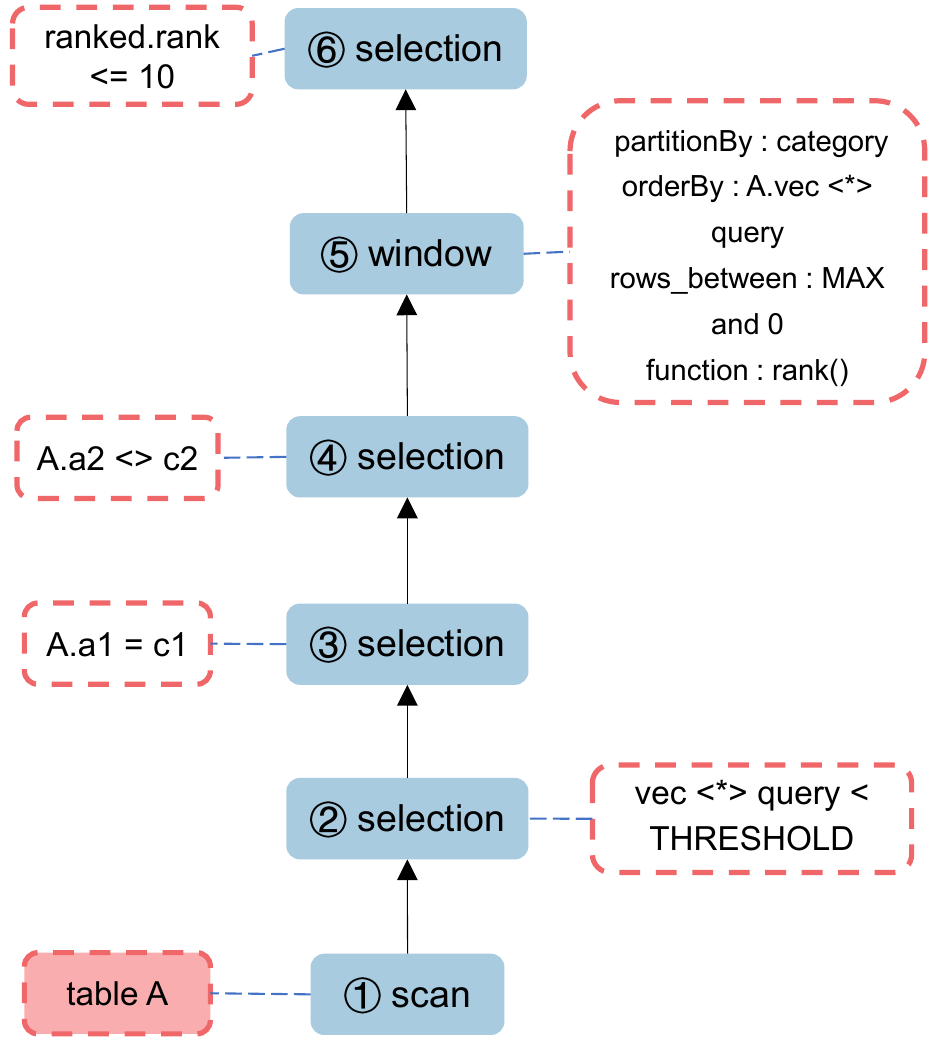}
         \caption{Unoptimized logical plan for Q5}
         \label{fig:uncategory_lplan}
     \end{subfigure}
     \begin{subfigure}[b]{0.24\textwidth}
    \captionsetup{font=scriptsize}
         \centering
         \includegraphics[width=1\textwidth]{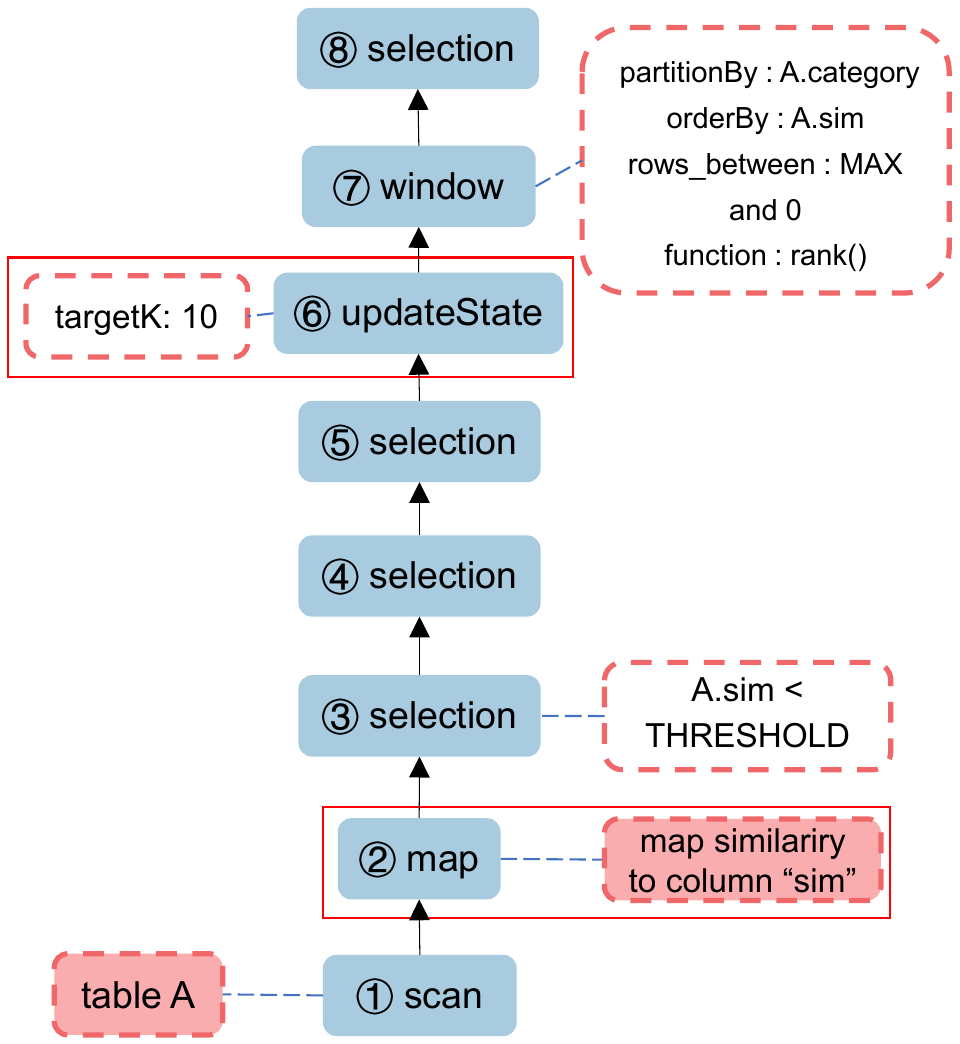}
        \caption{Optimized logical plan for Q5}
         \label{fig:category_lplan}
     \end{subfigure}
    \caption{Logical plan comparison for Q5}
\end{figure}

KNN-like queries are typically characterized by sorting data based on the similarity between items and a query vector, followed by selecting the top-\(K\) results. These queries adhere to a general pattern expressed as:
\[
\textit{orderBy}(\textit{D, distance}(\vec{v}_i, q)) \overset{R}{\rightarrow} \textit{topK}(R)
\]
First, the query incorporates an \textit{orderBy} operator that sorts the dataset \(D\) based on the similarity between the data items and the query vector \(q\), resulting in an intermediate dataset \(R\) where the most similar results are ranked at the top. Second, an additional operator is applied to select the top \(K\) results from this ordered dataset \(R\).
For instance, in the unoptimized logical plan corresponding to Q1 (Figure \ref{fig:untopk_lplan}), the \textit{orderBy} operator sorts data by similarity, while the \textit{limit} operator ensures that only the top \(K\) results are selected. Similarly, query plans such as those depicted in Figure \ref{fig:unknnjoin_lplan} and Figure \ref{fig:uncategory_lplan}, the \textit{window} operator employs \textit{orderBy} clause to rank tuples by similarity within partitions, followed by \textit{selection} operator to extract the top-\(K\) results. 
In the logical plans of such queries, the \textit{scan} operator can later be converted into an \textit{index scan} operator, enabling the utilization of the ANN index to accelerate hybrid query processing. In this case, the hybrid queries face the issue of redundant similarity computations between the \textit{scan} and \textit{orderBy} operators. 

To optimize such queries, CHASE introduces a \textit{map} operator, which maps the similarity scores obtained from the ANN index scan to a temporary column, denoted as "sim". Subsequently, the sorting field of the \textit{orderBy} operator is replaced with the "sim" column. After the plan rewriting, the logical plan, as shown in Figure \ref{fig:topk_lplan}, differs from the original logical plan depicted in Figure \ref{fig:untopk_lplan}. Specifically, in the rewritten plan, the sorting field of the \textit{orderBy} operator is no longer the expression (\textit{vec <*> query}). This change enables the \textit{orderBy} operator to directly leverage the similarity scores computed during the ANN index scan process rather than recalculating them. As a result, the result of the \textit{index scan} operator, passed through the \textit{map} operator, can be reused in the \textit{orderBy} operator, significantly reducing computational overhead.

\subsection{Rewriting Entity-Centric VKNN-SF Queries}
Entity-centric KNN queries require that each tuple in one table serves as a query to identify the top \(K\) related tuples in another table. Consider two tables: the query table \(T_q\) and the target table \(T_r\). Table \(T_q\) contains the entities to be queried, with each entity \(t_q \in T_q\) associated with a primary key \(pk_q\), while \(T_r\) holds the records related to the query entities, with each record \(t_r \in T_r\). These queries adhere to the following pattern:
\[
\begin{aligned}
&\textit{window}(T_q \Join T_r, \textit{partitionBy}(pk_q), \textit{size}: [0, MAX]) \\
&\xrightarrow{\textit{W}} \textit{orderBy\_per\_partition}(\textit{W, distance}(t_r, t_q)) \\
&\xrightarrow{\textit{R}} \textit{topK\_per\_partition}(R)
\end{aligned}
\]
The \textit{window} operator first partitions the data resulting from the join of the query table \(T_q\) and the target table \(T_r\) based on the primary key \(pk_q\) of the query table. This partitioning approach, which relies on the primary key, is essential to ensure that all tuples corresponding to the same entity are grouped together. If additional fields are included in the partitioning, tuples with the same primary key could be assigned to different partitions, thereby violating the integrity of the query semantics. Furthermore, the window size must span the entire partition, as indicated by \textit{size: [\text{0}, MAX]}, to ensure all candidate tuples are considered. If a subset of the partition is used, such as \textit{size: [\text{2}, 10]}, the search range for the nearest neighbors is altered, which hinders the effectiveness of the ANN index. The \textit{window} operator produces an intermediate result \(W\), which is subsequently ordered by the distance between the items in each partition and the query vector. This results in another intermediate dataset \(R\), from which the top-\(K\) results are selected from each partition.

When the hybrid query adheres to the aforementioned pattern, CHASE separates the \textit{orderBy} operator from the \textit{window} operator and inserts a \textit{limit} operator after the \textit{orderBy} operator, as demonstrated in the revised logical plan for Q4 (Figure \ref{fig:knnjoin_lplan}). This ensures that the \textit{scan}, \textit{orderBy}, and \textit{limit} operators are grouped into a single pipeline, whereas the \textit{join} operator serves as a pipeline breaker, leading to the formation of a VKNN-SF sub-query. As a consequence, during the selection of physical operators, the \textit{scan} operator can be optimized to utilize the ANN index, enabling it to efficiently retrieve the top \(K\) results for each tuple in the query table \(T_q\). 
In contrast, the original logical plan for Q4 (Figure \ref{fig:unknnjoin_lplan}) lacks the \textit{orderBy} and \textit{limit} operators within the same pipeline as the \textit{scan} operator. As a result, the query plan performs a brute-force search to filter a large set of candidate tuples from \(T_r\), which is computationally expensive and inefficient.
\subsection{Rewriting Category-Driven VKNN-SF Queries}
Category-Driven VKNN-SF queries differ from Entity-Centric VKNN-SF queries in both their partitioning method and filtering process. Category-Driven queries first filter \(T_r\) based on a distance threshold R1 when performing the join operation, and then partition by both \(pk_q\) and the categorization field \(c_r\), while Entity-Centric VKNN-SF queries join \(T_q\) and \(T_r\) and partition the results by the primary key \(pk_q\). As a result, the query pattern for Category-Driven VKNN-SF queries can be expressed as follows:
\[
\begin{aligned}
& \textit{window}(T_q \Join T_r \textit{ ON} \ \textit{distance}(t_r, t_q) \leq \text{R1}, \\
& \quad \quad \quad \quad \textit{partitionBy}(pk_q, c_r), \textit{size}: [0, MAX]) \\
& \xrightarrow {W} \textit{orderBy\_per\_partition}(\textit{W}, \textit{distance}(t_r, t_q)) \\
& \xrightarrow {R} \textit{topK\_per\_partition(R)} 
\end{aligned}
\]
The \textit{window} operator partitions data using a combination of the primary key from \(T_q\) and categorical field from \(T_r\). As with Entity-Centric VKNN-SF queries, the window size should span the entire partition to ensure all candidate tuples are considered, and the query must maintain KNN-like semantics to correctly identify top \(K\) results within each partition.

For hybrid queries that follow the Category-Driven VKNN-SF pattern, such as Q5 and Q6, CHASE introduces an \textit{updateState} operator before the \textit{window} operator to dynamically track tuples during query execution, as shown in Figure \ref{fig:category_lplan}. Unlike the original logical plan for Q5 (Figure \ref{fig:uncategory_lplan}), the introduction of the \textit{updateState} operator enables the system to monitor whether the minimal query range \( R_2 \) has been exceeded. When redundant computations are detected, the \textit{updateState} operator facilitates early termination of the scan operator, thereby preventing unnecessary evaluations of irrelevant tuples. By dynamically adjusting the query range, the system reduces computational costs while maintaining the accuracy of the query. The implementation detail of the \textit{updateState} operator is further elaborated in Section \ref{sec:physical_op_selection}.

\section{Physical Operators Optimization}\label{sec:physical_op_selection}
After rewriting the logical plan of hybrid query, CHASE selects pre-implemented physical operators from the operator library to optimize the execution of specific hybrid operators. 1) For DR-SF queries, we implement a range search algorithm for the \textit{index scan} operator, enabling the system to leverage ANN indices for processing DR-SF queries instead of relying on brute-force search. 2) Building on the optimized \textit{index scan} operator for DR-SF queries, we further implement the \textit{index scan} operator and the \textit{updateState} operator for Category-Driven VKNN-SF queries to efficiently track and manage the state of query execution. 3) Additionally, for KNN-like queries, we implement both the \textit{map} operator and optimized \textit{index scan} operator.

\subsection{Map Operator for KNN-like Queries} \label{sec:mapop}
To avoid recomputing the similarity scores already calculated during the index scan phase, CHASE introduces a \textit{map} operator for KNN-like queries, such as VKNN-SF queries, Entity-Centric VKNN-SF queries, and Category-Centric VKNN-SF queries. CHASE first modifies the output of the \textit{index scan} operator, which previously returned only the traversed tuples. Now, it also returns the computed similarity scores and passes them to the \textit{map} operator. The \textit{map} operator creates a new attribute for each tuple and adds the similarity score to this attribute, allowing subsequent \textit{sort} operator to directly use this precomputed result for sorting. Moreover, for queries Q1 and Q4, the \textit{scan} operator is replaced by the ANN Topk algorithm, which leverages the ANN index to quickly locate the neighborhood of the query vector and return the relevant tuples that satisfy the structural constraints.

\subsection{Index Scan Operator for DR-SF Queries}
For hybrid queries involving distance-based range conditions, such as Q2 and Q3, CHASE adopts the ANN-based range search algorithm employed by the VBASE system to optimize the \textit{index scan} operator for efficient range query execution. The core idea is to first use ANN search to quickly locate the neighboring nodes of the query vector, and then expand the search outward from these nodes to explore the query range centered around the query vector. By restricting the search to a refined subset of the dataset, this strategy obviates the need for exhaustive comparisons across all data using a sequential scan operator.

Algorithm \ref{alg:range_check} represents the execution process of the \textit{index scan} operator used for Q2 and Q3, where $hasInRange$ indicates whether the current query has entered the specified range, and $outRangeCounter$ tracks the number of consecutive out-of-range.
The algorithm begins by performing an ANN search to approach the query range \( R1 \) centered around the query \( q \), obtaining a similarity score \( sim \) \textit{(Line \ref{line:do_ann})}. If the similarity score is less than the threshold \( R1 \) \textit{(Line \ref{line:check_range})} and the query has not yet entered the range, it indicates that the search is still converging toward the target. Conversely, if the similarity score is less than \( R1 \) after the query has already entered the range \textit{(Lines \ref{line:check_range} - \ref{line:check_inrange})}, it implies that the search has moved beyond the query range. In this case, the algorithm checks whether the query has consecutively exceeded the range for \( N \) times \textit{(Line \ref{line:isalimit})}. If so, it concludes that all relevant data within the range has been traversed, and the search can terminate. Otherwise, the retrieval process continues \textit{(Line \ref{line:research})}. Finally, the algorithm returns the result tuple \( res \). In summary, the algorithm starts from an entry point in the ANN index, gradually converging toward the neighborhood of the query vector. From this neighborhood, it traverses the data within the query range until it consecutively exceeds the range \( N \) times.

\begin{algorithm}
\caption{Process of index scan operator for range search}
\label{alg:range_check}
\begin{algorithmic}[1]
\Statex \textbf{Input:} query $q$, similarity threshold $R1$, boolean reference $hasInRange$
\Statex \textbf{Output:} result tuple $res$

\State $outRangeCounter \gets 0$, $res \gets \emptyset$

\State $sim$, $res \gets \text{doANNSearch($q$)}$ \label{line:do_ann}
    
\If{$sim < R1$} \label{line:check_range}
    \If{$hasInRange$} \label{line:check_inrange}
        \State $outRangeCounter \gets outRangeCounter + 1$
        \State $needStop \gets \text{IsAboveN($outRangeCounter$)}$ \label{line:isalimit}
        \If{\textbf{not} $needStop$} 
            \text{ goto \ref{line:do_ann}} \label{line:research}
        \EndIf
    \EndIf
\Else
    \State $hasInRange \gets \textbf{TRUE}$ \label{line:done_search}
\EndIf
\State \Return $res$

\end{algorithmic}
\end{algorithm}

\subsection{UpdateState Operator for Category-Driven VKNN-SF Queries}
\begin{algorithm}
\caption{Execution process of updateState operator}
\label{alg:updaterecordtable}
\begin{algorithmic}[1]
\Statex \textbf{Input:} record table $T$, $tuple$ produced by other operators, $K$-value
\State $category, sim \gets \text{getInfo(tuple)}$ \label{line:get_tuple_info}

\If{$T.\text{lookup}(category) = \text{FALSE}$} \label{line:lookup_T}
    \State $T.restElements \gets T.restElements + 1$ \label{line:inc_counter}
\EndIf

\State $filteredK_c, queue_c \gets T.\text{lookupOrInsert}(category, sim)$ \label{line:lookup_insert_T}

\State $isMonotonic \gets \text{checkMonotonicity}(queue_c, sim)$ \label{line:check_Monotonicity}

\State $termination \gets filteredK_c \geq K$ \label{line:check_termination}

\If{$isMonotonic \land termination$} \label{line:check_need_stop}
    \State $T.restElements \gets T.restElements - 1$ \label{line:dec_counter}
\EndIf
\end{algorithmic}
\end{algorithm}

To minimize the query range required for handling Category-Driven VKNN-SF queries, CHASE employs the \textit{updateState} operator to dynamically maintain status information for each category during the query execution. This operator leverages the maintained information to assess the necessity of extending the current search range and directly influences the execution of the index scan operator. The \textit{index scan} operator can terminate its execution based on the status information maintained by the \textit{updateState} operator, effectively avoiding unnecessary range expansion and redundant computations.

Based on the execution process of the \textit{index scan} operator used for Q2 and Q3, CHASE introduces an additional check for the \textit{index scan} operator in Category-Driven VKNN-SF queries (Q5 and Q6). Specifically, in Algorithm \ref{alg:range_check}, a check is added before Line \ref{line:do_ann} as follows: \( if(T.restElements = 0) \ return \ res\). An extra input, the record table $T$, is introduced to store the status information of each category during the query process. Before performing the ANN search, the algorithm verifies whether there are any unprocessed categories in table $T$. If no unprocessed categories exist, the query can terminate early.

The \textit{updateState} operator introduced during the rewriting of the logical plan is responsible for updating the record table $T$. The execution process of the \textit{updateState} operator is illustrated in Algorithm \ref{alg:updaterecordtable}. 
The algorithm begins by extracting the category and similarity of the tuple produced by other operators \textit{(Line \ref{line:get_tuple_info})}. If the category does not exist in the record table, the count of unprocessed categories is incremented \textit{(Lines \ref{line:lookup_T} - \ref{line:inc_counter})}, indicating that a new category requiring processing is identified during the query. The table $T$ is implemented as a hash table, where each category is a key, mapping to associated search queues and a count of tuples that satisfy the predicate conditions during retrieval. Subsequently, the search queue for the current tuple's category is updated based on its similarity, and the monotonicity of the queue is verified to ensure that vector retrieval for that category is stabilizing. Additionally, the algorithm checks whether the number of retrieved tuples that satisfy the predicate conditions meets the specified size $K$ \textit{(Lines \ref{line:lookup_insert_T} - \ref{line:check_termination})}. If both the monotonicity and termination conditions are satisfied, the count of unprocessed categories is updated, indicating that the KNN query for that category have been completed within the current range. 

The \textit{updateState} operator and the \textit{index scan} operator work together to record the category of each tuple while traversing the data within the query range. When the number of categories in table \( T \) no longer changes, and the VKNN-SF query for each category has converged, it indicates that all categories within the current range have been processed and their respective VKNN-SF queries are complete. At this point, the query has been completed for Q5/Q6 within the current search range, which has reached the range \( R_2 \). If the search were to continue, the query range would need to be expanded to \( R_1 \), resulting in redundant computations.

\section{Code Generation} \label{sec:code_generation}
\begin{figure*}[ht]
    \centering
    \includegraphics[width=0.98\textwidth]{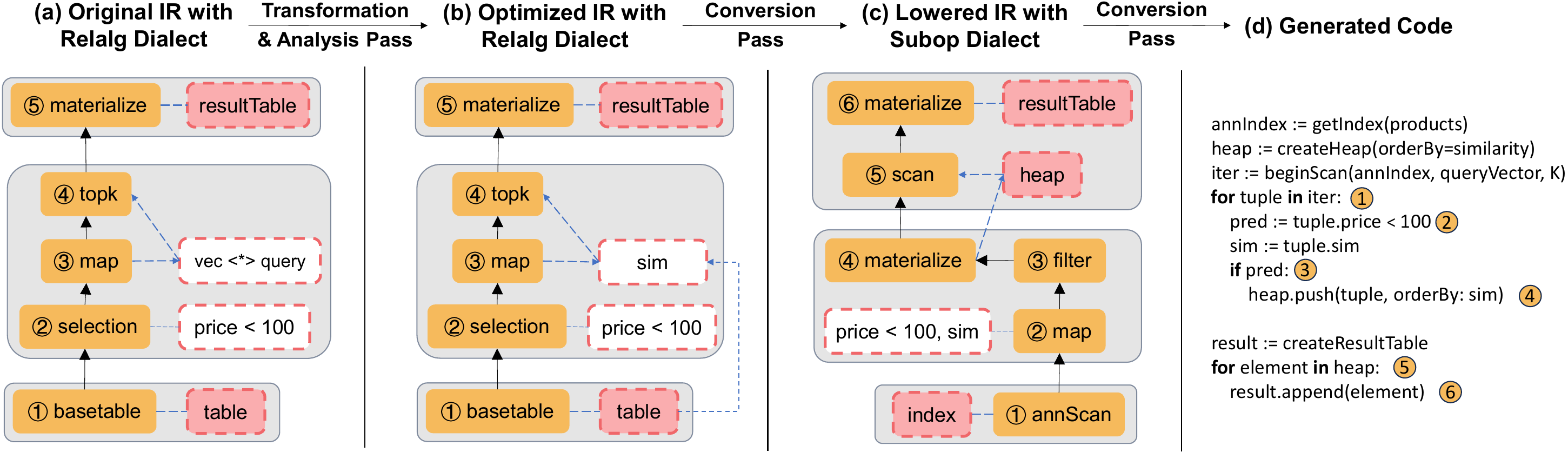}
    \caption{An example of code generation for Q1}
    \label{fig:lowering_example}
\end{figure*}
Code generation for hybrid queries aims to significantly improve the performance of query execution by translating high-level query plans into machine-level instructions that can be executed directly by the hardware. The primary goal is to reduce the overhead of traditional query execution engines, which often rely on high-level abstractions or interpreted code, leading to performance bottlenecks. By generating efficient machine code, systems can leverage the full computational power of modern hardware, optimizing for specific processor architectures and memory hierarchies. Furthermore, code generation enables more fine-grained optimizations, which are essential for maximizing performance in data-intensive tasks like hybrid querying.

The code generation in CHASE is implemented based on LingoDB \cite{jungmair2022designing, jungmair2023declarative}. LingoDB introduces five novel MLIR dialects designed to parse SQL queries into high-level MLIR modules. Subsequently, a series of optimization passes are applied to progressively transform the intermediate representation into low-level LLVM IR, which is eventually converted into machine code. CHASE extends the dialects and passes with minimal extensions.
Dialects simplify and optimize query compilation by introducing multi-level intermediate representations, allowing database query operations to be expressed at various abstraction levels. In CHASE, we extend several dialects with new operations: 1) In the \texttt{db} dialect, we introduce the \texttt{DenseVectorType<dim>} for vectors, as well as similarity computation operations like \texttt{L2Distance} and \texttt{InnerProduct}. 2) The \texttt{relalg} dialect is extended with the \textit{map} and \textit{updateState} operations, which are used during logical plan rewriting. 3) In the \texttt{subop} dialect, the index type \texttt{ExternalANNIndexType} is introduced, indicating that the object of the index scan is an ANN index. Additionally, the \texttt{annScan} operation is added to traverse the ANN index. 
Passes are structured techniques employed by the LLVM compiler for analyzing, optimizing, and transforming compilation objects, such as IR. CHASE defines a series of passes to rewrite the logical plan, select physical operators, and generate machine code. 

Taking Q1 as an example, CHASE parses the SQL query to produce the initial IR, as illustrated in Figure \ref{fig:lowering_example}(a). The original IR consists of operations from the relalg dialect.
Next, CHASE analyzes the initial IR using predefined analysis passes to determine whether it conforms to the patterns mentioned in Section \ref{sec:query_rewriting}. For this initial IR, the \texttt{relalg.topk} operation sorts the tuples based on their similarity to the query vector and selects the top \( K \) results, which aligns with the KNN-like query pattern. Therefore, CHASE applies a transformation pass to rewrite the IR and convert it into an optimized form, as illustrated in Figure \ref{fig:lowering_example}(b). Since the \textit{map} operation has already been introduced in the initial IR, CHASE only needs to modify the fields of the \texttt{relalg.map} operation to the "sim" attribute, thereby storing the similarity scores obtained from the index traversal of the \texttt{relalg.basetable} operation into the "sim" attribute. This entire process is equivalent to logical plan rewriting.

Then, the relalg dialect in the optimized IR is further lowered to the subop dialect through conversion passes, which is equivalent to selecting physical operators for the logical operators. The IR after lowering is illustrated in Figure \ref{fig:lowering_example}(c).
Compared to the optimized IR with relalg dialect, the \texttt{relalg.basetable} operation is converted into the \tikz[baseline=(char.base)]{\node[draw,circle,inner sep=1.0pt] (char) {1};} \texttt{subop.annScan} operation, which utilizes the Topk algorithm of the ANN index to find the K nearest neighbors of the query vector. Moreover, the \texttt{relalg.map} operation is transformed into the \tikz[baseline=(char.base)]{\node[draw,circle,inner sep=1.0pt] (char) {2};} \texttt{subop.map} operation, which stores the similarity score into the "sim" attribute while also checking whether the "price" attribute of the current tuple is less than 100.

Eventually, the query plan is compiled into machine code. During this process, several general passes are also applied to the IR, such as common subexpression elimination (CSE), dead code elimination, and constant folding, further enhancing query processing optimization.

\section{Evaluation} \label{sec:evaluation}
In this section, we evaluate CHASE and compare it with relational databases that support hybrid queries. Through experimental results, we conduct a detailed analysis of CHASE's performance and accuracy in hybrid queries involving relational and vector data, aiming to highlight its advantages in handling hybrid query scenarios.

\subsection{Benchmark}
\textbf{Dataset. }We evaluate the system using the publicly available dataset LAION-400-MILLION OPEN DATASET, which is divided into two subsets, laion1m and queries. We select the LAION dataset for our evaluation because it combines relational data with embedding vectors, making it particularly suitable for hybrid query processing. The laion1m subset consists of 1 million data entries, while the queries subset contains 100 entries. The data between the two tables is mutually exclusive. 
The schemas of both tables are illustrated in Table \ref{tab:table_schema}, where sample\_id serves as a unique identifier to distinguish each tuple. The url field points to the link address of the stored image files, and the text field contains textual descriptions associated with the images. The nsfw field employs the CLIP model to assess whether an image contains adult or inappropriate content, yielding results classified as "UNLIKELY," "UNSURE," or "NSFW." Additionally, the similarity field represents the cosine similarity value between the text and image embeddings, with higher values indicating stronger relevance. The width and height fields denote the dimensions of the image embeddings, while the vec field represents the embedding for each image, with a dimensionality of 512.

\captionsetup[table]{skip=2pt} 
\begin{table}[htbp]
\caption{Schema of Laion table}
        \label{tab:table_schema}
    \begin{threeparttable}
        \begin{tabular}{|c|c|c|}
        \hline
        \textbf{Column}& \textbf{Type}& \textbf{Example} \\
        \hline
        sample\_id \tnote{*} & int8  & 10869007972 \\
        \hline
        url & text  & "http://example.com/a.jpg"  \\
        \hline
        text & text  & "Santa Claus Suit Costume" \\
        \hline
        height & int4  & 250 \\
        \hline
        width & int4  & 320 \\
        \hline
        nsfw & text  & "UNLIKELY" \\
        \hline
        similarity & float4  & 0.346 \\
        \hline
        vec & float4$[512]$  & $[0.02,0.08,-0.01,-0.04...]$ \\
        \hline
        \end{tabular}
        \begin{tablenotes}
            \footnotesize 
            \item[*] PRIMARY KEY. 
        \end{tablenotes}
    \end{threeparttable}
\end{table}

\textbf{Queries. }We evaluate the three types of hybrid queries mentioned in Section \ref{sec:hybrid_query} using the six SQL query templates illustrated in Figure \ref{fig:sql_code_listings}. For each SQL query, we design six levels of selectivity to enhance query complexity, specifically: 1, 0.9, 0.7, 0.5, 0.3, and 0.03. The selectivity is defined as \(\text{selectivity} = \frac{\text{Number of tuples satisfying the predicate}}{\text{Total number of tuples}}\).
Notably, when selectivity equals 1, it indicates that no relational data is involved in the query. To generate queries with specific selectivities, we employ a quantile-based analytical approach. Specifically, we extract the target column from the dataset and calculated its corresponding quantiles. These quantiles provide the proportion of tuples in the entire dataset that fall below these values, thereby aligning the ratio of tuples satisfying the predicate to the total number of tuples as expected. For Q1 and Q4, we set \( K \) to 50; for Q5 and Q6, the \( K \) value for each category is set to 10. In the case of Q2 and Q3, we establish a threshold for the query range at 0.8, ensuring that, on average, each tuple within the range can match approximately 120 tuples.

\textbf{Metric. }We adopt \textbf{recall} as the measure of accuracy, defined as \(\text{recall} = \frac{|X \cap G|}{|G|}\), where \( G \) represents the set of actual neighbor samples, and \( X \) denotes the set of samples obtained through ANN search. A higher recall indicates that the system can effectively capture more true samples. Additionally, we introduce \textbf{execution time} as a key metric for performance evaluation. The measurement of execution time excludes planning time, data loading time, and compilation time, ensuring that only the actual computation time during input data processing is recorded.

\subsection{Experiment Setup}
The experimental setup utilizes a machine with a dual-socket configuration, housing two Intel(R) Xeon(R) Gold 5218 CPUs, each operating at a base frequency of 2.30 GHz. The system comprises 64 virtual CPUs (vCPUs), with 16 cores per socket and 2 threads per core. The CPU features a maximum frequency of 3.90 GHz and includes cache sizes of 1 MiB for L1 cache, 32 MiB for L2 cache, and 44 MiB for L3 cache. 

\textbf{Baseline systems.} We select VBASE, PASE, and pgvector database systems as baselines, all of which are built on PostgreSQL to handle hybrid queries involving relational and vector data. Additionally, we incorporate LingoDB as a baseline to highlight the performance differences between databases based on data-centric code generation and traditional query execution models. Since the native LingoDB does not support vector operations, we extend it to include this functionality, naming the modified version LingoDB-V. To ensure fairness in comparison, given that VBASE is developed based on PostgreSQL 13.4, we standardize all systems to this version, configure the \textit{shared\_buffers} parameter to 5GB, and utilize the \textit{pg\_prewarm} extension to preload relevant tables and indices into memory, thereby ensuring that query processing across all systems occurs in memory. Furthermore, both LingoDB-V and CHASE utilize an 8-thread default configuration for query processing. Similarly, the \textit{max\_parallel\_workers} parameter for VBASE, PASE, and pgvector is also set to 8 to ensure consistent parallelism across all systems.

\textbf{Index Settings}. With the exception of LingoDB-V, all systems use HNSW as the ANN index. For index construction, the maximum number of neighbors \(M\) is set to 16, with \(ef\_construction\) set to 200 and  \(ef\_search\) set to 48. The metric employed is the inner product.

\subsection{Performance Comparison}

Next, we discuss the detailed evaluation results for each hybrid query. Some baseline systems are unable to utilize ANN indices when processing hybrid queries, which results in a recall rate of 1. For these systems, we represent their recall as '-' in the tables, as shown in Table \ref{tab:topk_result}, indicating that they cannot leverage ANN indices.

\begin{table*}[h]
    \centering
    \caption{Average execute time(ms) and recall for Q1}
    \begin{tabular}{|l|cc|cc|cc|cc|cc|cc|}
        \hline
        \multirow{2}{*}{DBName} & \multicolumn{2}{c|}{Selectivity = 1} & \multicolumn{2}{c|}{Selectivity = 0.9} & \multicolumn{2}{c|}{Selectivity = 0.7} & \multicolumn{2}{c|}{Selectivity = 0.5} & \multicolumn{2}{c|}{Selectivity = 0.3} & \multicolumn{2}{c|}{Selectivity = 0.03} \\ 
            \cline{2-13}
            & time & recall & time & recall & time & recall & time & recall & time & recall & time & recall \\ 
            \hline
            lingodb-v    & 126.49 & - & 53.94 & - & 47.44 & - & 39.42 & - & 28.27 & - & \textbf{8.82} & - \\
                        
            pgvector   & 5.11 & 0.98 & 6.05 & 0.98 & 8.14 & 0.98 & 9.58 & 0.98 & 13.18 & 0.98 & 33.72 & 0.52 \\
            
            pase       & 2.22 & 0.98 & 2.30 & 0.98 & 2.65 & 0.98 & 2.99 & 0.98 & 4.19 & 0.98 & 18.12 & 0.98 \\ 
            vbase      & 2.31 & 0.98 & 2.41 & 0.98 & 2.82 & 0.98 & 3.54 & 0.98 & 4.88 & 1 & 22.15 & 1 \\
            
            chase  & \textbf{1.90} & \textbf{0.98} & \textbf{2.08} & \textbf{0.98} & \textbf{2.4} & \textbf{0.98} & \textbf{2.79} & \textbf{0.98} & \textbf{3.66} & \textbf{1} & 14.88 & 1 \\
            \hline
    \end{tabular}
    \label{tab:topk_result}
\end{table*}

\begin{table*}[h]
    \centering
    \caption{Average execute time(ms) and recall for Q2}
    \begin{tabular}{|l|cc|cc|cc|cc|cc|cc|}
        \hline
        \multirow{2}{*}{DBName} & \multicolumn{2}{c|}{Selectivity = 1} & \multicolumn{2}{c|}{Selectivity = 0.9} & \multicolumn{2}{c|}{Selectivity = 0.7} & \multicolumn{2}{c|}{Selectivity = 0.5} & \multicolumn{2}{c|}{Selectivity = 0.3} & \multicolumn{2}{c|}{Selectivity = 0.03} \\ 
            \cline{2-13}
            & time & recall & time & recall & time & recall & time & recall & time & recall & time & recall \\ 
            \hline
            lingodb-v    & 590 & - & 596 & - & 586 & - & 596 & - & 585 & - & 585 & - \\
            pgvector   & 834 & - & 843 & - & 734 & - & 687 & - & 619 & - & 567 & - \\
            pase       & 2,209 & - & 2,096 & - & 1,772 & - & 1,513 & - & 1,244 & - & 909 & - \\ 
            vbase      & 13.58 & 0.9991 & 12.93 & 0.9990 & 14.96 & 0.9987 & 12.15 & 1 & 11.99 & 1 & 11.8 & 1 \\
            \textbf{chase}  & \textbf{9.1} & \textbf{0.9991} & \textbf{9.26} & \textbf{0.9990} & \textbf{10.1} & \textbf{0.9987} & \textbf{9.94} & \textbf{1} & \textbf{9.05} & \textbf{1} & \textbf{8.98} & \textbf{1} \\
            \hline
    \end{tabular}
    \label{tab:range_result}
\end{table*}

\subsubsection{Q1: VKNN-SF Queries}
Table \ref{tab:topk_result} illustrates the performance and recall of various systems under six selectivity levels for executing Q1. In the recall experiments, all systems supporting HNSW (excluding LingoDB-V) achieve high recall at a selectivity of 1, where relational data is excluded. For pgvector and PASE, when the selectivity decreases while keeping the \textit{ef\_search} parameter constant, the recall also decreases. To align the recall across all systems, we manually adjust the \textit{ef\_search} parameter for pgvector and PASE. In contrast, CHASE adopts the approach from VBASE, with its ability to dynamically adjust search queue lengths, maintaining recall consistently above 0.98, demonstrating the advantage of adaptive queue management.
In terms of execution time, CHASE demonstrates significant improvements due to optimizations in the \textit{map} operator and query compilation. For instance, at a selectivity of 1, CHASE outperforms VBASE by 17\%, despite both systems employing the same retrieval algorithm. As selectivity decreases, the performance gap between CHASE and VBASE grows from 17\% to 33\%. 

In addition, we evaluate the impact of query compilation optimizations by analyzing hardware metrics such as branch misses, L1 data cache misses, and the number of instructions across varying selectivity levels. As shown in Table \ref{tab:miss_analysis}, CHASE consistently shows a significant reduction in branch misses, cache misses, and the number of instructions executed compared to VBASE, highlighting the improved efficiency by machine code generation. 



\begin{table*}[ht]
\centering
\caption{Branching and cache locality in Q1}

\begin{tabular}{@{}lcccccccccccc@{}}
\toprule
\textbf{Metric} & \multicolumn{2}{c}{\textbf{Selectivity = 1}} & \multicolumn{2}{c}{\textbf{Selectivity = 0.9}} & \multicolumn{2}{c}{\textbf{Selectivity = 0.7}} & \multicolumn{2}{c}{\textbf{Selectivity = 0.5}} & \multicolumn{2}{c}{\textbf{Selectivity = 0.3}} & \multicolumn{2}{c}{\textbf{Selectivity = 0.03}} \\ 
\cmidrule(lr){2-3} \cmidrule(lr){4-5} \cmidrule(lr){6-7} \cmidrule(lr){8-9} \cmidrule(lr){10-11} \cmidrule(lr){12-13}
                 & chase & vbase & chase & vbase & chase & vbase & chase & vbase & chase & vbase & chase & vbase \\ \midrule
Branches    & 1.32M & 929M & 1.23M & 928M & 1.44M & 926M & 1.93M & 928M & 2.78M & 929M & 11.4M & 962M \\ 
Branches misses      & 4.97K & 2.16M & 5.15K & 2.23M & 6.27K & 2.43M & 8.11K & 2.38M & 10.9K & 2.37M & 40.7K & 2.39M \\ 
Branch miss rate      & 0.38\% & 0.23\% & 0.42\% & 0.24\% & 0.44\% & 0.26\% & 0.42\% & 0.26\% & 0.39\% & 0.25\% & 0.36\% & 0.25\% \\ 
Instructions      & 9.75M & 4.14B & 10.0M & 4.13B & 12.1M & 4.14B & 15.4M & 4.13B & 18.1M & 4.14B & 85.5M & 4.33B \\ 
L1-D reference      & 5.24M & 1.11B & 5.11M & 1.12B & 6.27M & 1.12B & 8.18M & 1.12B & 11.1M & 1.12B & 46.2M & 1.17B \\ 
L1-D misses      & 78.2K & 90.6M & 70.4K & 90.1M & 79.7K & 90.1M & 100K & 89.7M & 134K & 89.9M & 564K & 94.8M \\ 
L1-D miss rate      & 1.49\% & 8.14\% & 1.37\% & 8.07\% & 1.27\% & 8.06\% & 1.22\% & 8.03\% & 1.20\% & 8.05\% & 1.21\% & 8.09\% \\ 

\bottomrule
\end{tabular}
\label{tab:miss_analysis}
\end{table*}

\begin{table}[ht]
    \centering
    \caption{Average execute time(ms) and recall for Q3}
    \begin{tabular}{|l|>{\centering}m{7mm}m{7mm}<{\centering}|>{\centering}m{6mm}m{7mm}<{\centering}|>{\centering}m{6mm}m{7mm}<{\centering}|}
        \hline
        \multirow{2}{*}{DBName} & \multicolumn{2}{c|}{Selectivity = 1} & \multicolumn{2}{c|}{Selectivity = 0.5} & \multicolumn{2}{c|}{Selectivity = 0.03} \\ 
            \cline{2-7}
            & time & recall & time & recall & time & recall \\ 
            \hline
            lingodb-v    & 2.63$\times 10^3$ & - & 2.78$\times 10^3$ & - & 191 & - \\

            pgvector   & 2.35$\times 10^4$ & - & 1.18$\times 10^4$ & - & 1.32$\times 10^3$ & - \\
            
            pase       & 1.16$\times 10^5$ & - & 5.75$\times 10^4$ & - & 4.28$\times 10^3$ & - \\ 
            
            vbase      & 195 & 0.93 & 173 & 0.94 & 176 & 0.95 \\
            
            chase  & \textbf{62} & \textbf{0.93} & \textbf{61} & \textbf{0.94} & \textbf{62} & \textbf{0.95} \\
            \hline
    \end{tabular}
    \label{tab:distancejoin_qps}
\end{table}

\subsubsection{Q2-3: DR-SF Queries}
Table \ref{tab:range_result} presents the performance of databases in handling Q2. It is evident that databases supporting the RangeSearch interface of the ANN index, such as VBASE and CHASE, significantly outperform other baseline systems in terms of execution time. Compared to brute-force search, the number of similarity computations is reduced from 1 million to just 4,800. Additionally, due to the advantages of machine code generation, CHASE consistently delivers the best performance across all selectivity levels. CHASE outperforms the best-performing baseline system, VBASE, by 24\% to 33\% in terms of performance, while maintaining the same recall rate.

Notably, only pgvector and PASE show a reduction in query execution time as selectivity decreases. This can be attributed to the fact that, during query execution, the system first filters the relational data and only performs vector similarity calculation on the data that satisfies the structured constraints. As selectivity decreases, the amount of data that meets the structured constraints reduces, leading to fewer similarity calculations and thus decreasing the overall query execution time. In contrast, VBASE, LingoDB-V, and CHASE follow a different query processing strategy: they first filter tuples based on similarity requirements and then apply relational data constraints. This approach results in a constant number of similarity calculations as selectivity decreases. Additionally, the experimental dataset consists of 512-dimensional vectors, and the computation of high-dimensional vector similarity constitutes a significant portion of the query execution time. Therefore, the query execution time for VBASE, LingoDB-V, and CHASE do not exhibit significant changes as selectivity decreases. 

When executing Q3, PASE and pgvector incur high materialization costs, leading to significantly higher query execution time compared to LingoDB-V, as illustrated in Table \ref{tab:distancejoin_qps}. LingoDB-V benefits from its ability to fully utilize 8 threads, significantly improving its performance. In contrast, although both PASE and pgvector are configured with a parallelism of 8 processes, they fail to fully utilize all available processes during execution, and their process-level communication, which is more time-consuming than thread-level communication, further increases execution time compared to LingoDB-V.
CHASE inherits the multi-threading characteristic of LingoDB-V, allowing it to maintain optimal performance across all selectivity conditions, with recall \(\geq\) 0.93. At the same recall level and with the same number of similarity computations, CHASE is approximately 64\% faster than VBASE.

\subsubsection{Q4: Entity-Centric VKNN-SF Queries}

\begin{table}[h]
    \centering
    \caption{Average execute time(ms) and recall for Q4}
    \begin{tabular}{|l|>{\centering}m{7mm}m{7mm}<{\centering}|>{\centering}m{6mm}m{7mm}<{\centering}|>{\centering}m{6mm}m{7mm}<{\centering}|}
        \hline
        \multirow{2}{*}{DBName} & \multicolumn{2}{c|}{Selectivity = 1} & \multicolumn{2}{c|}{Selectivity = 0.5} & \multicolumn{2}{c|}{Selectivity = 0.03} \\ 
            \cline{2-7}
            & time & recall & time & recall & time & recall \\ 
            \hline
            lingodb-v    & 8.67$\times 10^3$ & - & 5.40$\times 10^3$ & - & 392 & - \\

            pgvector   & 1.95$\times 10^5$ & - & 9.38$\times 10^4$ & - & 5.83$\times 10^3$ & - \\
            
            pase       & 3.34$\times 10^5$ & - & 1.63$\times 10^5$ & - & 1.02$\times 10^4$ & - \\ 
            
            vbase      & 3.33$\times 10^5$ & - & 1.62$\times 10^5$ & - & 1.02$\times 10^4$ & - \\
            
            chase  & \textbf{26} & \textbf{0.95} & \textbf{30} & \textbf{0.95} & \textbf{215} & \textbf{0.97} \\
            \hline
    \end{tabular}
    \label{tab:knn_qps}
\end{table}
CHASE demonstrates exceptional performance in Q4, completing the query in just 26 milliseconds at selectivity = 1 while maintaining a recall rate of 0.95. This significant performance improvement is primarily attributed to the use of the ANN index, which reduces the number of similarity computations from 100 million to only 94,000.
As a result, CHASE is 7,500\(\times\) faster than pgvector and 330\(\times\) faster than LingoDB-V, as shown in Table \ref{tab:knn_qps}. Even when selectivity drops to 0.03, CHASE retains its performance advantage, completing the query in only 215 milliseconds, still 45\% faster than the fastest baseline system, LingoDB-V.
In contrast, baseline systems cannot leverage ANN indices while handling KNN-join queries, forcing them to perform linear scans of both tables. Moreover, systems such as pgvector, PASE, and VBASE require materialization during the join operation, which leads to higher execution costs due to the additional memory required to store intermediate results. Specifically, VBASE and PASE store vectors as float8 arrays, which incurs higher memory costs during materialization, resulting in slower execution time compared to pgvector. In contrast, LingoDB-V avoids the need to materialize the table multiple times and employs a hash-based approach to execute the Partition BY clause, reducing its query execution time.


\subsubsection{Q5-6: Category-Driven VKNN-SF Queries}
\begin{figure}[h]
    \centering
    \begin{subfigure}[b]{0.23\textwidth}
    \captionsetup{font=scriptsize}
         \centering
         \includegraphics[width=1\textwidth]{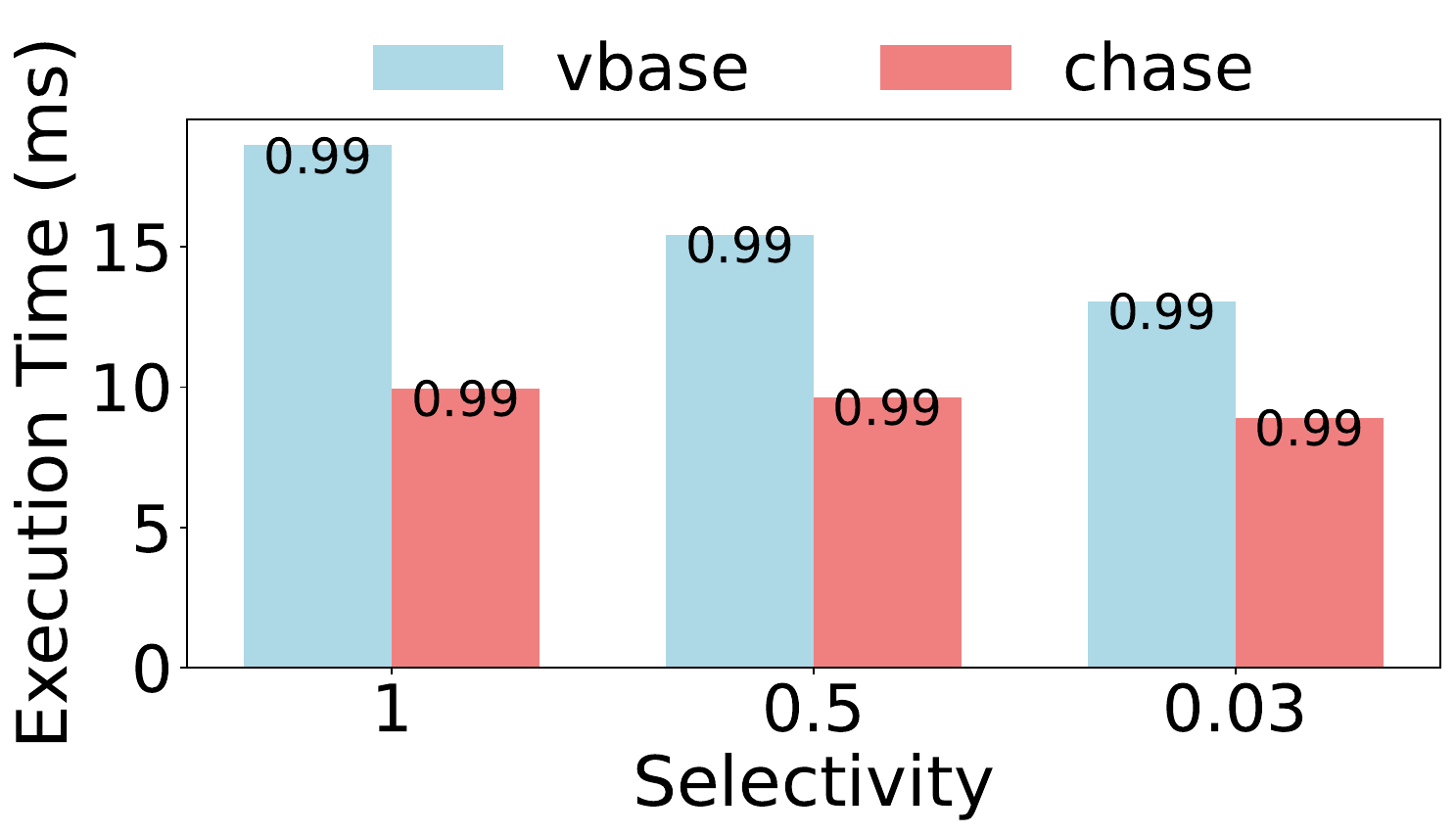}
         \caption{Performance comparison of Q5}
         \label{fig:categorybased_result}
     \end{subfigure}
     \begin{subfigure}[b]{0.23\textwidth}
    \captionsetup{font=scriptsize}
         \centering
         \includegraphics[width=1\textwidth]{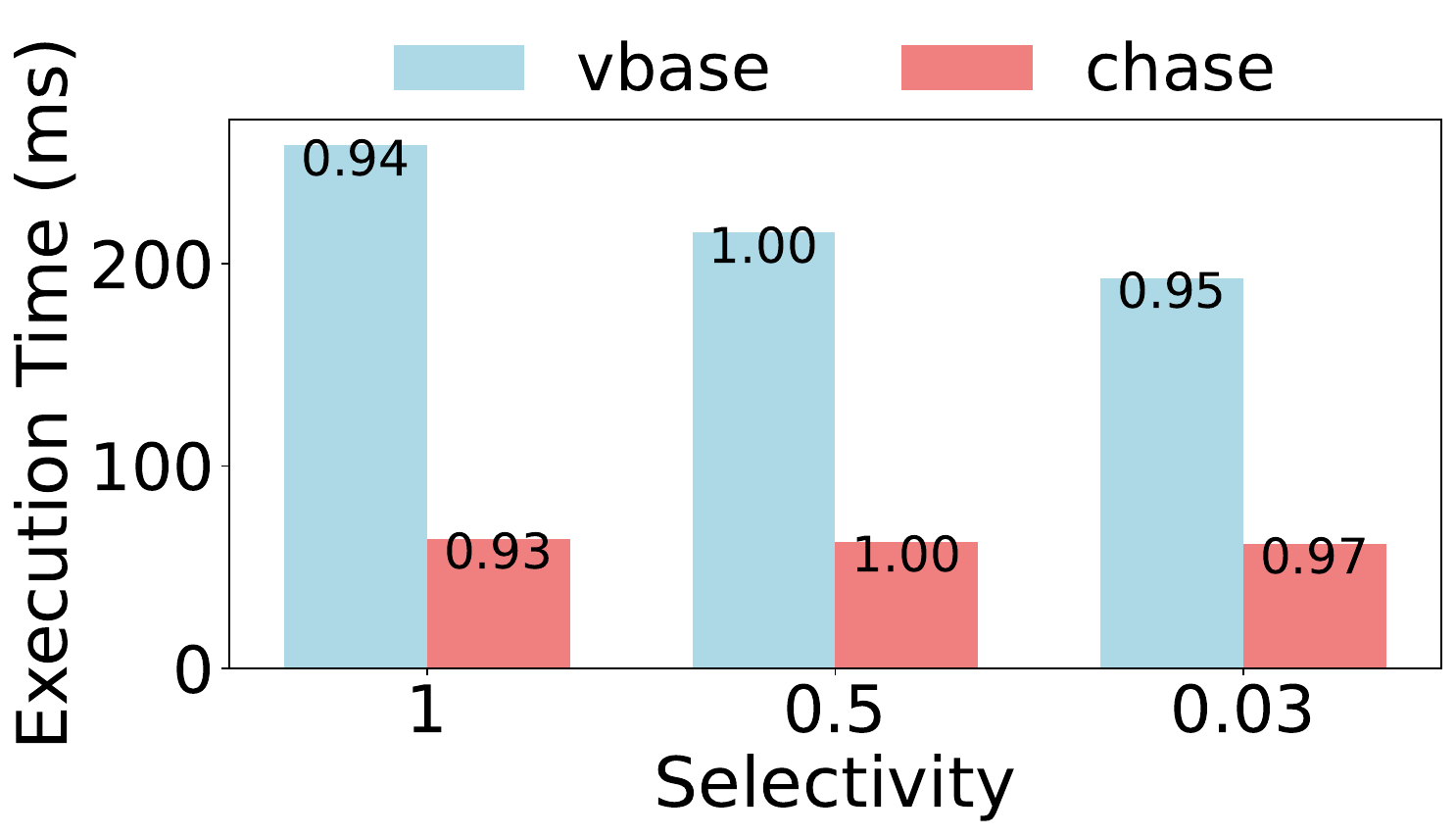}
        \caption{Performance comparison of Q6}
         \label{fig:categoryjoin_result}
     \end{subfigure}
    \caption{Performance comparison of Q5 and Q6}
    \label{fig:category_results}
\end{figure}
\begin{figure}[h]
    \centering
    \begin{subfigure}[b]{0.23\textwidth}
    \captionsetup{font=scriptsize}
         \centering
         \includegraphics[width=1\textwidth]{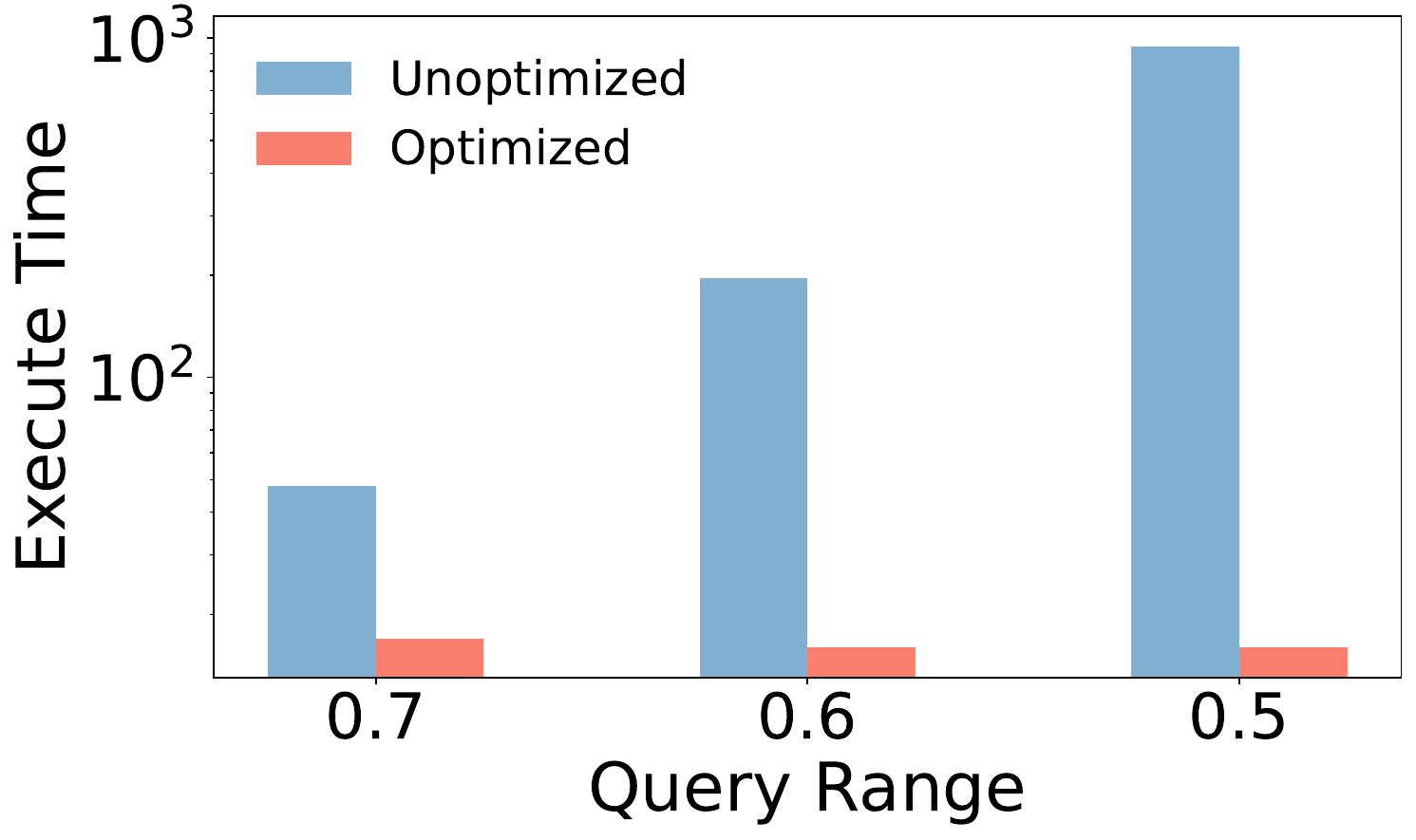}
         \caption{Execution time of Q5}
         \label{fig:categoryrange_qps}
     \end{subfigure}
     \begin{subfigure}[b]{0.23\textwidth}
    \captionsetup{font=scriptsize}
         \centering
         \includegraphics[width=1\textwidth]{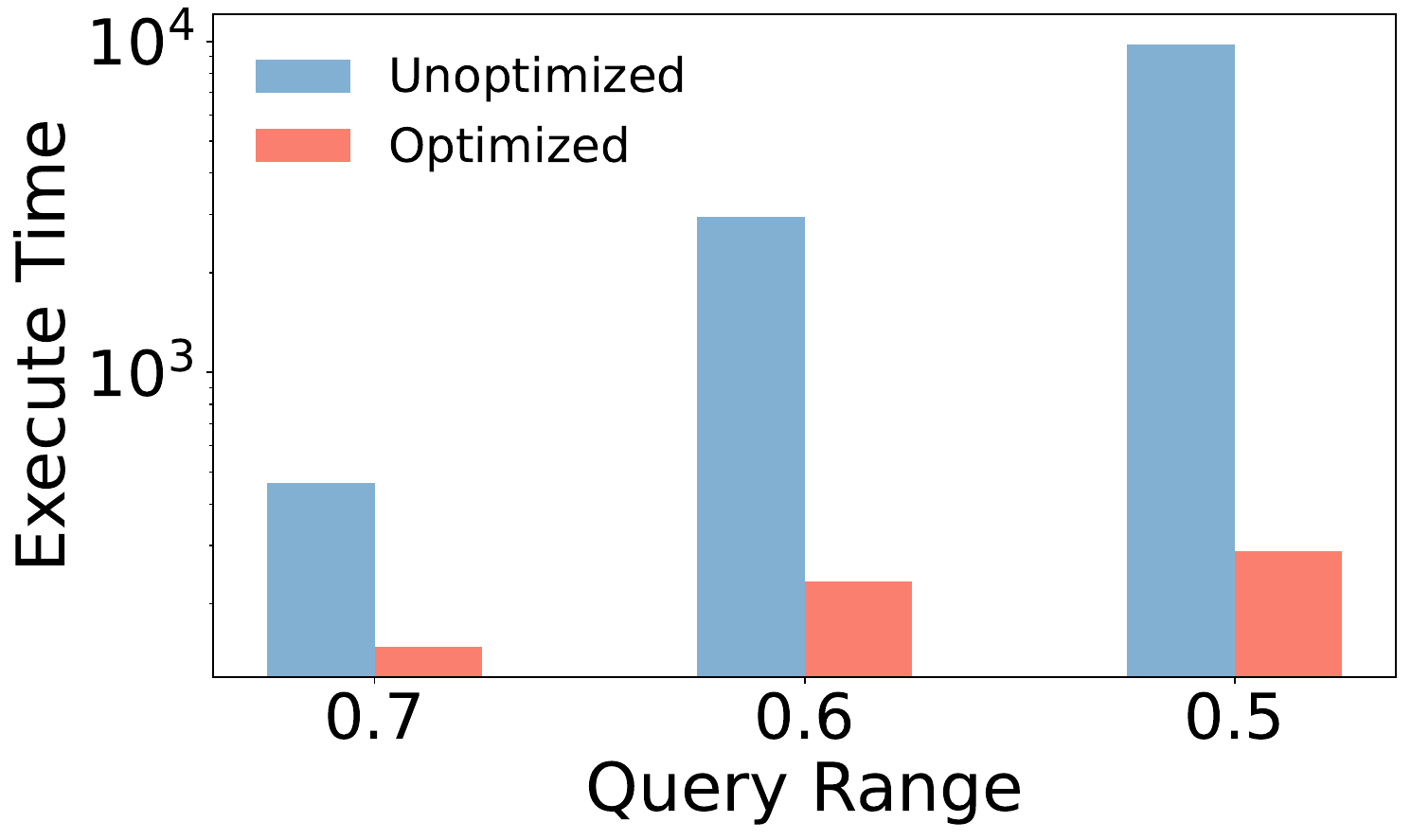}
        \caption{Execution time of Q6}
         \label{fig:categoryjoinrange_qps}
     \end{subfigure}
    \caption{Impact of updateState operator on Q5 and Q6}
\end{figure}
Since entity-Centric VKNN-SF queries are similar to DR-SF queries and exhibit similar results, we select only VBASE and CHASE as example systems in Figure \ref{fig:category_results}. Overall, for category-based partition queries (Q5), CHASE is 33\% to 46\% faster than VBASE. For category-join queries (Q6), CHASE performs 3.1\(\times\) to 4.04\(\times\) faster than VBASE. Under the experimental setup, the query range for Q5 and Q6 is set to 0.8. At this range, both VBASE and CHASE perform the same number of similarity computations during index traversal, which demonstrates that even with a relatively small query range, CHASE achieves a performance improvement without relying on \textit{updateState} operator.

To validate the effectiveness of \textit{updateState} operator, we additionally configure three similarity thresholds of 0.7, 0.6, and 0.5, with selectivity set to 1. In this context, a smaller similarity threshold corresponds to a larger query range, increasing the number of data to be processed. As illustrated in Figure \ref{fig:categoryrange_qps}, without the \textit{updateState} operator in the query plan, the execution time significantly increases as the similarity threshold decreases, increasing from 48 ms to 946 ms. In contrast, the query plan with the \textit{updateState} operator stabilizes the execution time at 16 ms, confirming the effectiveness of the \textit{updateState} operator.
It is noteworthy that, as shown in Figure \ref{fig:categoryjoinrange_qps}, as the query range expands, the execution time of CHASE for Q6 does not remain as stable as it does for Q5, even with the introduction of the \textit{updateState} operator. This is because the \textit{index scan} operator terminates the query early if the specified range is not reached after a certain number of retrieval attempts. Consequently, some subqueries may not be terminated as early as they were when the query range was smaller, leading to an increase in overall execution time. However, the increase in execution time is not significant compared to the execution time of the unoptimized query plan.

\section{Conclusion} \label{sec:conclusion}
We introduce CHASE, a query engine built with native support for executing hybrid queries on structured and unstructured data. We investigate a broad class of hybrid queries from modern applications and analyze their overheads when designing CHASE. CHASE optimizes the overall performance through logical plan rewriting, physical operator optimizations, and machine code generation, enabling efficient integration of vector similarity search and structured data filtering in the relational database. Evaluations on real-world datasets demonstrate significant performance gains over existing systems, demonstrating CHASE as a robust solution for handling diverse and complex hybrid queries. 

\bibliographystyle{ACM-Reference-Format}
\bibliography{sample}

\end{document}